\documentclass{emulateapj}
\usepackage{natbib,graphicx,psfig}

\def \Msun {\, \rm M_\odot}
\def \be {\begin{equation}}
\def \ee {\end{equation}}

\begin{document}

\title{Are the Magellanic Clouds on their First Passage about the Milky Way?}
\shorttitle{Orbital evolution of the Clouds}
\author{Gurtina Besla\altaffilmark{1}, Nitya Kallivayalil\altaffilmark{1}, Lars Hernquist\altaffilmark{1}, Brant Robertson\altaffilmark{2,3,4}, T.J. Cox\altaffilmark{1}, Roeland P. van der Marel\altaffilmark{5}, Charles Alcock\altaffilmark{1}}
\altaffiltext{1}{Harvard-Smithsonian Center for Astrophysics, 60 Garden Street,
Cambridge, MA 02138}
\altaffiltext{2}{Kavli Institute for Cosmological Physics and Department of Astronomy and Astrophysics, University of Chicago, 933 East 56th Street, Chicago, IL 60637}
\altaffiltext{3}{Enrico Fermi Insitute, 5640 South Ellis Avenue, Chicago, IL 60637}
\altaffiltext{4}{Spitzer Fellow}
\altaffiltext{5}{Space Telescope Science Institute, 3700 San Martin Drive, Baltimore, MD 21218}
\email{gbesla@cfa.harvard.edu}

\begin{abstract}
Recent proper motion measurements of the Large and Small Magellanic
Clouds (LMC and SMC, respectively) by \citet{Nitya1, Nitya2} suggest
that the 3D velocities of the Clouds are substantially higher
($\sim$100 km/s) than previously estimated and now approach the escape
velocity of the Milky Way (MW).  Previous studies have also assumed
that the Milky Way can be adequately modeled as an isothermal sphere
to large distances. Here we re-examine the orbital history of the
Clouds using the new velocities and a $\Lambda$CDM-motivated MW model
with virial mass $M_{vir}=10^{12}\Msun$ (e.g. \citet{Klypin}). We
conclude that the L/SMC are either currently on their first passage
about the MW or, if the MW can be accurately modeled by an isothermal
sphere to distances\ $\gtrsim 200$ kpc (i.e., $M_{vir}>2\times
10^{12}\Msun$), that their orbital period and apogalacticon distance
must be a factor of two larger than previously estimated, increasing
to 3 Gyr and 200 kpc, respectively.  A first passage scenario is
consistent with the fact that the LMC and SMC appear to be outliers
when compared to other satellite galaxies of the MW: they are
irregular in appearance and are moving faster. We discuss the
implications of this orbital analysis for our understanding of the
star formation history, the nature of the warp in the MW disk and the
origin of the Magellanic Stream (MS), a band of HI gas trailing the
LMC and SMC that extends $\sim$100 degrees across the
sky. Specifically, as a consequence of the new orbital history of the
Clouds, the origin of the MS may not be explainable by current tidal
and ram pressure stripping models.
\end{abstract}

\keywords{galaxies: interactions --- galaxies: kinematics and dynamics --- galaxies: evolution ---  Galaxy: structure --- Magellanic Clouds }

\section{Introduction}
\label{sec:intro}

Accurate estimates of the current positions and velocities of the
Large and Small Magellanic Clouds (LMC and SMC, respectively) are
crucial parameters for modeling their past interactions with the Milky
Way (MW). Recent proper motion measurements with the {\it Hubble Space
Telescope (HST)} of the LMC and SMC by \citet[][hereafter {\bf K1} and
{\bf K2}]{Nitya1,Nitya2} imply a substantial increase ($\sim100$ km/s)
in the estimate of the current 3D galactocentric velocities of the
Clouds. This paper explores the consequences of these new measurements
for our understanding of the global dynamics of the MW-LMC-SMC system.

Constraints on the orbital history of the Clouds can be obtained from
the structural and kinematic properties of the Magellanic Stream
(MS). The MS is a band of HI gas structured in filaments and clumps
extending $\sim$100$\degr$ across the sky, nearly along a great circle
beginning at the Clouds and passing through the south galactic pole
\citep{Bruens2005, Putman2003}. From radial velocity measurements of
the HI gas, the Clouds are known to be leading the MS
\citep{Putman}. Also, since the potential well of the Clouds is
shallow compared to that of the MW, the orbit of the Clouds should be
similar to that of the MS \citep{Fich}. Thus, the Clouds are believed
to be currently following a nearly polar orbit of counter-clockwise
sense as seen from the Sun.

Previous studies of the orbital evolution of the Clouds and the
formation of the MS (e.g. \citet[][hereafter \bf{MF80}]{MF},
\citet[][\bf{LL82}]{Lin}, \citet[][\bf{GSF94}]{GSF},
\citet[][\bf{HR94}]{Heller}, \citet{Moore}, \citet{Lin95},
\citet[][\bf{GN96}]{GN}, \citet{Bekki}, \citet{Yoshizawa},
\citet{Connors}, \citet[][\bf{M05}]{Mastro}) generally follow the
prescription introduced by MF80: the differential equations of motion
for both the LMC and SMC are integrated numerically, allowing the
position and velocity of the Clouds to be followed backwards in time.
Because of the large uncertainties in previous proper motion estimates
for the Clouds 
{\bf vdM02}]{vanderMarel}, the listed authors chose orbital parameters
that best reproduced the properties of the MS under the assumption
that the LMC and SMC form a binary system that has been in a slowly
decaying orbit about the MW for roughly a Hubble time ($t_{\rm H}$).

The existence of a common envelope of diffuse HI gas surrounding the
Clouds supports the assumption of binarity for at least some time in
the past \citep{Mathewson}. However, the only constraint on the
orbital period of this binary system about the MW is the length of
time required to form the MS. This timescale is largely dependent on
the proposed formation mechanisms: generally, theories invoke some
combination of ram pressure and tidal stripping, implicitly assuming
that the Clouds have undergone multiple pericentric passages. These
arguments will be revisited critically in $\S$\ref{subsec:Imp}.

Recently, K1 and K2 have measured the systemic proper motions of the
LMC and SMC by using the Advanced Camera for Surveys (ACS) on {\it HST}
 to track the LMC(SMC)'s motion relative to 21(5) background QSOs
discovered from their optical variability in the MACHO database.  This
has made it possible to determine the proper motion of the LMC to
better than 5$\%$ accuracy: $\mu_W = -2.03 \pm 0.08 {\rm mas/yr}$, 
$\mu_N = 0.44 \pm 0.05 {\rm mas/yr}$ (K1). They also obtain a four-fold
improvement in accuracy over previous measurements of the SMC's proper
motion: $\mu_W = -1.16 \pm 0.18$ mas/yr, $\mu_N = -1.17 \pm 0.18$
mas/yr (K2). The corresponding mean 3D space velocities and one-sigma
errors for the LMC are listed in row four of
Table~\ref{Table:summary}. A lower limit on the interaction time
between the LMC/SMC binary and MW is thus $\sim$250 Myr, which
corresponds to the amount of time it takes the LMC to travel the
length of the MS, assuming that the MS extends along a circle of
radius 55 kpc and that the LMC is moving at $v=378$ km/s (the mean 3D
velocity determined by K1).

Previous estimates of the LMC's proper motions are summarized in
Table~\ref{Table:PMsummary}. The K1 proper motions differ
substantially from those used in the theoretical models of
GN96. However, only the west component of the K1 values differs
markedly from the weighted average of previous measurements determined
by vdM02. In this analysis we focus mainly on the {\it HST}
measurements, although we also consider the vdM02 proper motion values
as a point of reference. We have not explicitly included the recent
\citet{Pedreros} results in our analysis as this work is essentially a
refinement of the previous \citet{Pedreros02} measurements, which are
already included in the vdM02 weighted average.

The significant error reduction in the {\it HST} proper motion
measurements has also limited the range of plausible current orbital
parameters. Specifically, the 4$\sigma$ lower limit on the LMC's
present 3D velocity is now $\sim310$ km/s, which is larger than the
GN96 estimate of 297 km/s. Correspondingly, if we model the MW as an
isothermal sphere, the predicted apogalacticon distance reached by the
LMC on its last orbit would be
roughly twice as large are previously predicted (see
$\S$~\ref{sec:iso} and Table~\ref{Table:summary}). This change owes to
the substantial increase in the estimate of the LMC's tangential
velocity component (i.e. $\mu_W$). The large tangential velocity
($v_{tan}$) and relatively small radial velocity ($v_{rad}$) imply
that the Clouds are likely either near apogalacticon or
perigalacticon. Earlier works, such as that of \citet{Lynden-Bell} and
\citet{Fujimoto}, assumed that the Clouds were near apogalacticon;
however, the debate was effectively settled by early proper motion
measurements which unequivocally showed that $v_{tan} > 220$ km/s
(i.e. larger than roughly the maximum circular velocity).
 The centrifugal force is thus larger than the gravitational force acting on the LMC, meaning the LMC is moving toward larger radii ($dr/dt>0$) and so must be at perigalacticon. 

\begin{deluxetable*}{ccccccccc}
\tabletypesize{\scriptsize}
\tablecaption{Summary of LMC Orbital Parameters Adopted in Previous Studies}
\tablewidth{0pt}
\tablehead{
\colhead{Work} & \colhead{3D $v$ (x,y,z) (km/s)\tablenotemark{a}} & \colhead{$|v|$ (km/s)} & \colhead{$v_{\rm tan}$ (km/s)} & \colhead{$v_{\rm rad}$ (km/s)} & \colhead{$r$ (x,y,z) (kpc)} & \colhead{T (Gyr)}\tablenotemark{b} & \colhead{Peri (kpc)} & \colhead{Apo (kpc)} }
\startdata
MF80 & (233.7,-13.1,252.4)\tablenotemark{c} & 344 & 340 & 92 & (42.9,-2.4,-28.3) & 1.5 & 50 & 110 \\ 
GSF94, GN96 & (-5,-226, 194)\tablenotemark{c} & 297 & 287 & 82 & (-1.0,-40.8,-26.8) & 1.5 & 45 & 120  \\ 
HR94 & (-10.06,-287.09,229.73)\tablenotemark{c} & 367.83 & 351.81 & 107.37 & (-0.85,-40.85,-27.95) & 2.5 & 46.3 & 180 \\  
vdM02 & (-56 $\pm$ 36, -219 $\pm$ 23, & 293 $\pm$ 39 &  281 $\pm$ 41 & 84 $\pm$ 7 & (-0.8, -41.5, -26.9) & 2 & 45 & 110 \\ 
 & 186 $\pm$ 35)\tablenotemark{d} & & & & & & & \\
M05\tablenotemark{e} & (-4.3,-182.45,169.8)\tablenotemark{d} & 249.3 & 237.9 & 74.4 & (0,-43.9,-25.04) & 2 & 45 & 115 \\
K1 Mean & (-86 $\pm$ 12, -268 $\pm$ 11, & 378 $\pm 18$ & 367 $\pm$ 18 & 89 $\pm$ 4 & (-0.8,-41.5,-26.9) & 3  &  50 & 220 \\
 & 252 $\pm$ 16)\tablenotemark{d} &  &  &  &  &  &  &  \\
K2 Fig. 12  & (-91, -250, 220)\tablenotemark{d} & 345 & 333 & 92 & (-0.8,-41.5,-26.9) & 2 & 50 & 150 \\
\enddata
\tablenotetext{a}{All positions and velocities are measured in Galactocentric coordinates.}
\tablenotetext{b}{T is the orbital period} 
\tablenotetext{c}{These velocities are predictions from theoretical models.}
\tablenotetext{d}{These velocities are derived from observations of the LMC's proper motions using the distance moduli of 18.50 $\pm$ 0.1 \citep{Freedman}. 1$\sigma$ errors are quoted.}
\tablenotetext{e}{Parameters for M05 were obtained via private communication (2007).} 
\label{Table:summary}
\end{deluxetable*}

\begin{deluxetable*}{cccc}
\tabletypesize{\scriptsize}
\tablecaption{Summary of LMC Proper Motions Estimates}
\tablewidth{0pt}
\tablehead{
\colhead{Work} & \colhead{$\mu_W$ (mas/yr)} & \colhead{$\mu_N$ (mas/yr)} & \colhead{Method}}
\startdata
GSF94,GN96 & -1.72 & 0.12 & MS model \\
HR94\tablenotemark{a} & -2.0 & 0.16 & MS model  \\ 
vdM02 & -1.68 $\pm$ 0.16 & 0.34 $\pm$ 0.16 & Compilation of pre-2002 measurements\tablenotemark{b} \\
K1 & -2.03 $\pm$ 0.08 & 0.44 $\pm$ 0.05 & {\it HST} measurements \\ 
\enddata
\tablenotetext{a}{These are our estimates of the LMC's proper motions corresponding to the 3D velocities used by HR94, determined using the distance moduli from \citet{Freedman}.}
\tablenotetext{b}{The vdM02 value is the weighted average of the ground-based and {\it Hipparcos} measurements by: \citet{Kroupa94, Jones, Kroupa97, Pedreros02, Drake}. }
\label{Table:PMsummary}
\end{deluxetable*}

Yet, a tangential velocity comparable to that measured by K1 (367 km/s) has been considered by just three studies before: MF80, LL82 and HR94 ($v_{tan}$ = 340, 370 and 350 km/s, respectively). They determined the orbital parameters of the LMC by requiring that its projected orbit on the sky traced the current location of the MS (see $\S$\ref{subsec:MS}) and that the Clouds have been bound to one another for $\sim10^{10}$ Gyr. However, GSF94 made the same assumptions and yet determined a significantly lower tangential velocity ($v_{tan}$ = 287 km/s). The requirement that the LMC's projected orbit traces the MS only uniquely determines $\mu_N$; there exists a substantial degeneracy in possible $\mu_W$ values when a 3D orbit is viewed in projection on the plane of the sky (see $\S$\ref{subsec:MS}). Since it is $\mu_W$ that ultimately determines the tangential velocity component, it becomes relatively clear how these groups could make the same assumptions and obtain such different estimates for $v_{tan}$.  For example, although the values of $\mu_N$ determined by the theoretical models of HR94, LL82 and GSF94 are roughly the same (Table~\ref{Table:PMsummary}), their estimates for $\mu_W$ are quite different. GSF94 also matched the line-of-sight radial velocities along their orbit to the available HI data in the MS in determining their proper motions; but, this assumption is dependent on the advocated formation mechanism of the MS, which is still a subject of debate (see $\S$\ref{subsubsec:stream}). The discrepancy between the $\mu_W$ components in these studies ultimately owes to the details of the MW models and choice of perigalacticon, which strongly affects the lifetime of the binary system. 


From Table~\ref{Table:summary}, only the HR94 orbital parameters are consistent with those of the K1 orbits in an isothermal sphere model. However, as the apogalacticon distance increases, the validity of an isothermal sphere model for the MW is questionable: simulated DM halos don't behave like isothermal spheres, especially at large radii.
 Furthermore, the corresponding increase in orbital period limits the effectiveness of ram pressure and tidal stripping, which are two popular mechanisms for the formation of the MS. 

In this paper we revisit this classic orbital mechanics problem, adopting 3D velocities constrained by the recent proper motion measurements of K1(K2) for the LMC(SMC) and a more detailed MW model. In $\S$\ref{sec:iso} we first reconsider the isothermal sphere model to highlight the substantial change in the LMC's orbital history implied by the new measurements. In $\S$\ref{sec:orbit} we consider instead a cosmologically-motivated model and describe our methodology and model parameters:
our fiducial MW model is consistent with model A$_1$ of \citet[][hereafter \bf{KZS02}]{Klypin} (virial mass $M_{vir}= 10^{12} \Msun$) and known observational constraints. We focus our analysis on the LMC alone as the LMC:SMC mass ratio of $\sim$10:1 precludes the SMC from being a major determinant in the LMC's orbital history.
In $\S$\ref{sec:results} the error space of the new proper motion measurements is searched to test all allowed orbital histories for the LMC. 
For our fiducial MW model, the new mean 3D velocity is roughly the escape velocity at the LMC's current location (50 kpc). Consequently, our computed orbital histories differ substantially from the predictions of all previous studies:
 we find that the orbit of the LMC is best described as highly eccentric and approaching parabolic, implying that the LMC is on its first passage about the MW. We further show that orbits computed with either our fiducial model or the isothermal sphere model are co-located when projected on the plane of the sky, but deviate from the current location of the MS. Moreover, this deviation exists even if the vdM02 proper motion measurements are used. 
 In $\S$\ref{subsec:model} our results are shown to be robust to changes in our fiducial model parameters.
 We also consider the impact of the SMC on the robustness of our results in $\S$\ref{subsec:SMC}. We discuss the reliability of the new proper motion measurements in $\S$\ref{subsec:PM} and comment on the likelihood of a first passage scenario in $\S$\ref{subsec:first}. The implications of the new orbital history for the star formation history of the LMC, the production of the warp in the MW disk and the formation of the MS are discussed in $\S$\ref{subsec:Imp}. We conclude in $\S$\ref{sec:conc}.

\section{Earlier Models}
\label{sec:iso}

The present galactocentric 3D space, tangential and radial velocities, positions, past orbital period, perigalacticon and apogalacticon distances for the LMC as determined by MF80, GSF94, GN96, HR94, vdM02 and M05 are listed in Table \ref{Table:summary}. These works serve as characteristic examples of the range of orbital parameters previously considered for the Magellanic system. 

With the exception of M05, the listed authors all modeled the MW as a
singular isothermal sphere and accounted for the effects of dynamical
friction using the Chandrasekhar formula (\citet{BT} equation 7-18) with a
constant Coulomb logarithm of ln$\Lambda$ in the range $1-3$. 
M05 modeled the dark matter halo of the MW using an
NFW profile \citep{NFW} with concentration parameter $c=11$, virial
radius $R_{vir}= 200$ kpc and a virial mass of $M_{vir}$ =
$1.1\times10^{12} \Msun$. However, they also considered a
substantially lower 3D velocity for the LMC (250 km/s). 
The listed authors were able to find solutions where the
Clouds follow quasi-periodic orbits that slowly decay owing to
dynamical friction/tidal stripping as the Clouds move through the dark
matter halo of the MW.

The 3D space velocities and 1$\sigma$ errors for the LMC as determined
by K1 are listed in row four of Table~\ref{Table:summary}. K2 follow
the MF80 prescription to search for solutions within their proper
motion error distribution in which the LMC and SMC maintain a stable
binary system for roughly a Hubble time: a characteristic set of these
solutions
is listed in row five of Table~\ref{Table:summary} (see Figure 12 in
K2). Although their analysis illustrates that the new proper motion
measurements allow for orbital histories that are roughly consistent
with previous models, the apogalacticon distances have increased
substantially: the LMC now reaches distances $>$150 kpc on its
previous orbital passage, compared to distances of $\sim$100 kpc
determined by previous authors. 

\begin{figure}[t]
\begin{center}
\includegraphics[scale=0.43]{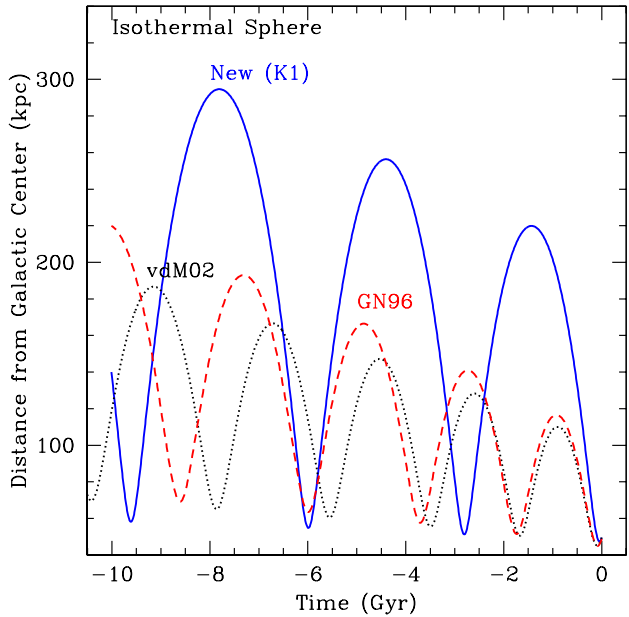}
\caption{Galactocentric radial distance plotted as a function of time for the mean K1 (solid, blue line), GN96 (dashed, red line) and vdM02 (dotted, black line) values. The MW dark matter halo is modeled as an isothermal sphere and the effect of dynamical friction is approximated using the Chandrasekhar formula with Coulomb logarithm ln$\Lambda =3$.  Even in this idealized case, the apogalacticon distance reached using the mean K1 values is $\sim$ two times larger than the GN96 or vdM02 results and the orbital period has increased to $\sim$ three Gyr.}
\label{fig:Isothermal}
\end{center}
\end{figure}

Using the mean K1 3D velocities (row four of
Table~\ref{Table:summary}), the apogalacticon distance increases to
220 kpc and the orbital period to three Gyr. This contrasts starkly with 
the orbit corresponding to the vdM02 averaged value, which yields results 
similar to those of GN96.
The full orbital history of the LMC estimated using the K1 mean values and 
those of vdM02 and GN96 are
plotted for comparison in Figure~\ref{fig:Isothermal} (blue solid line, black 
dotted line and red dashed line, respectively). Only the orbital parameters of
HR94 are consistent with the new picture.

Although the K1 mean values still imply that the LMC has completed
several pericentric passages about the MW within a Hubble time, the
increased orbital period and apogalacticon distance present
substantial challenges to current tidal and ram pressure stripping
models for the formation of the Magellanic stream (see
$\S$\ref{subsubsec:stream}).
Moreover, it is unclear that an isothermal sphere model is appropriate
at distances $\gtrsim$200 kpc.  This is certainly true over long
timescales since dark matter halos evolve over time 
\citep{Wechsler}
For example, the work of \citet{Wechsler} suggests that the formation
 time of a MW-type halo (defined as the time when the halo was half as
 massive as today) is $\sim$8 Gyr ago.
 If the MW halo were smaller and less massive in
the past, the orbital energy of a satellite would increase as
we integrate its orbit backwards in time \citep{Penarrubia}. The
apogalacticon distance is thus expected to increase more dramatically
in the past than shown in Figure~\ref{fig:Isothermal}. In light of
this, we opt for a more detailed MW model as described in the next
section.

\section{Methodology \& Model Parameters}
\label{sec:orbit}

Using the current 3D velocity and position of the LMC the differential equations of motion (\ref{eqmotion}) can be solved numerically, allowing the position and velocity of the LMC to be followed backwards in time (see MF80).  We consider only the gravitational influence of the MW and dynamical friction owing to the passage of the LMC through the dark matter halo of the MW: 

\begin{equation}
\ddot{\rm \bf {r}} = \frac{\partial}{\partial \bf {r}} \phi_{\rm MW}
\left (|\bf {r}|\right)  + \frac{\rm \bf {F}_{\rm DF}}{\rm M_{\rm {LMC}}},
\label{eqmotion}
\end{equation} 
where $M_{\rm {LMC}}$ is the mass of the LMC, $\bf {r}$ is its position vector, $\phi_{\rm MW}$ is the potential of the MW and $\bf {\rm F}_{\rm DF}$ is the dynamical friction term. We trace the orbital history of the LMC from equation (\ref{eqmotion}) using the symplectic leapfrog integration scheme outlined in \citet{GADGET}. 

Instead of an isothermal sphere model, we describe the MW as a static, axi-symmetric, 4-component model, consisting of a dark matter (DM) halo, Miyamoto-Nagai disk \citep{Miyamoto}, Hernquist bulge \citep{Hernquist} and a halo of hot gas in hydrostatic equilibrium with the DM. 
\begin{equation}
\Phi_{MW} = \Phi_{DM Halo} + \Phi_{Hot Halo} + \Phi_{disk} + \Phi_{bulge}
\label{eq:Phi}
\end{equation}
The DM halo is initially modeled as an NFW halo. We then use CONTRA, a publically available code written by Oleg Gnedin \citet{Oleg}, to model the adiabatic contraction of the NFW halo in response to the slow growth of an exponential disk. CONTRA follows either the \citet{Blumenthal} algorithm, which assumes spherical symmetry, homologous contraction and circular particle orbits, or a modified algorithm developed by \citep{Oleg} that attempts to account for the high orbital eccentricities of particles predicted in hierarchical structure formation scenarios. We use only the \citet{Blumenthal} algorithm for simplicity: as we demonstrate in $\S$\ref{subsec:mass}, the model parameters do not substantially change our results. The Miyamoto-Nagai profile is the monopole expansion of the exponential disk profile and is found to be an adequate approximation. 

Our fiducial model is consistent with both model A$_1$ of KZS02 ($M_{vir} = 10^{12}M_\sun$) and known observational constraints. Characteristic parameters for the fiducial model are summarized in column two of Table~\ref{Table:model}. The third column, labeled High Mass, will be discussed in $\S$~\ref{subsec:mass}. In all cases we adopt a flat $\Lambda$CDM cosmology with parameters consistent with KZS02: $h=0.7$, $\Omega_{M} =0.3$ and baryon fraction $f_b=\Omega_b/\Omega_M = 0.1$. 
 According to the spherical top hat model \citep{Gunn}, the virial radius ($R_{vir}$) is defined as the radius where the average mass enclosed ($\rho_M$) equals the virial density ($\rho_{vir} = \Delta_{vir}\rho_M$). We follow the analysis of KZS02 and take $\Delta_{vir}\approx 340$ for our cosmological model. Assuming a static MW halo, $R_{vir}$ is defined as:
\begin{equation}
R_{vir}=206h^{-1}{\rm kpc}\left(\frac{\Delta_{vir}\Omega_{M}}{97.2} \right)^{-1/3}\left(\frac{M_{vir}}{10^{12}h^{-1}M_\sun}\right)^{1/3}
\label{eq:Rvir}
\end{equation} 


\begin{figure}[t]
\begin{center}
\includegraphics[scale=0.43]{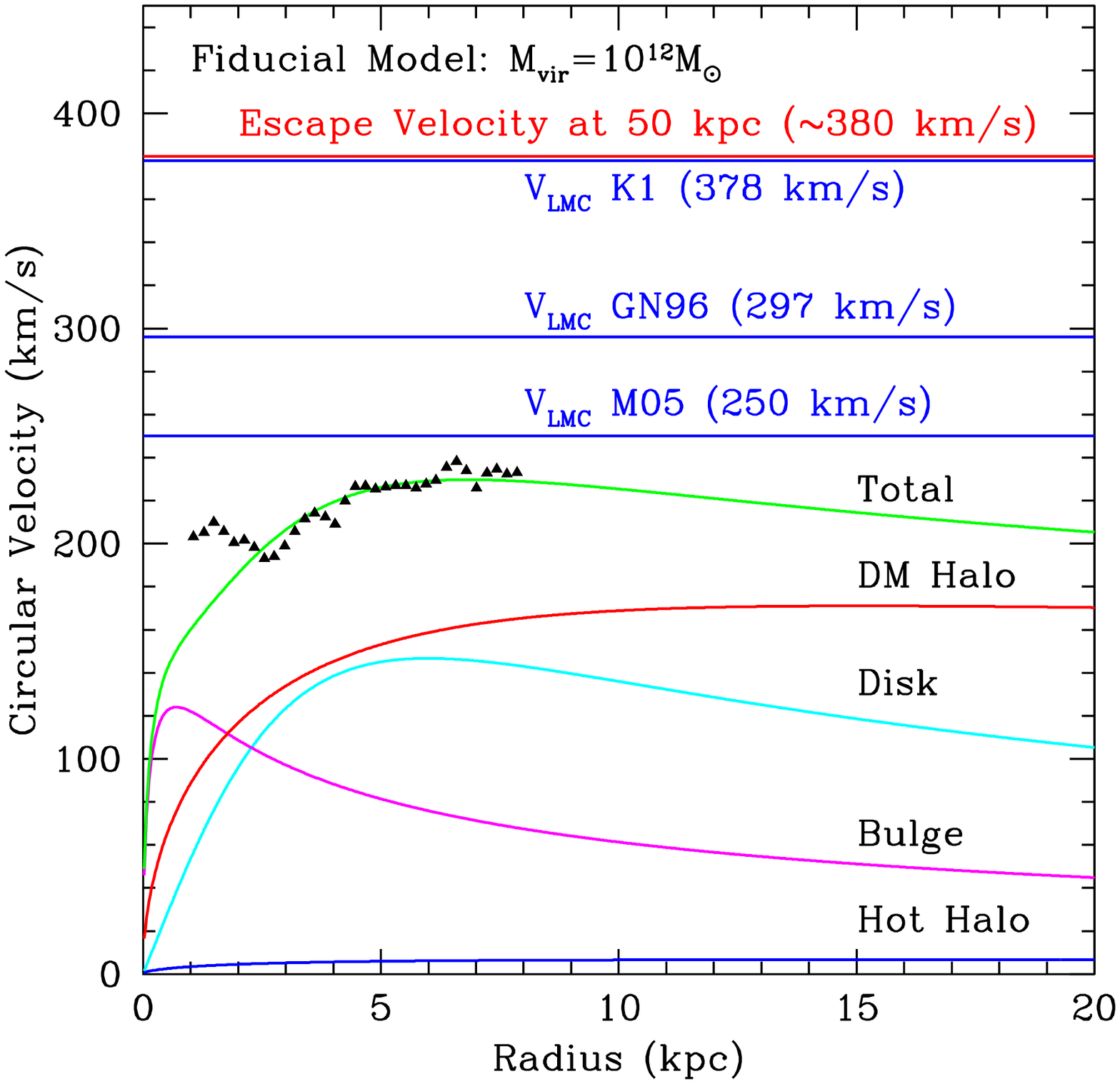}
\caption{The rotational velocity profile for our fiducial MW model is shown within 20 kpc. The circular velocity profile for each model component is also plotted individually. Horizontal blue lines indicate the 3D velocity estimates of GN96, M05 and K1. The K1 value is close to the red line indicating the escape velocity at 50 kpc. The triangles represent the HI measurements of \citet{Knapp} and were used to constrain the rotation curve within the solar circle (8 kpc).}
\label{fig:Fiducial}
\end{center}
\end{figure}

The resulting rotation curve within 20 kpc is plotted in Figure~\ref{fig:Fiducial}. The blue horizontal lines indicate the 3D velocity estimates of M05, GN96 and K1. Notice that the M05 velocity is close to the circular velocity whereas that of K1 is $\sim$ the escape velocity at 50 kpc; i.e. at the current location of the LMC.  Triangles represent observational constraints from HI measurements by \citet{Knapp}, where the solar radius is assumed to be eight kpc. Known observational constraints and predictions from the MW models of KZS02, \citet{DehnBinn} and \citet{Smith} are compared to our models in Table~\ref{Table:Constraints}: all values are shown to be consistent with our fiducial model. 

\begin{deluxetable*}{ccc}
\tabletypesize{\scriptsize}
\tablecaption{Model Parameters}
\tablewidth{0pt}
\tablehead{
\colhead{Parameter} & \colhead{Fiducial}\tablenotemark{a} & \colhead{High Mass\tablenotemark{b}}}
\startdata
Virial Mass $M_{vir}$ ($M_\sun$) & $10^{12}$ & $2\times10^{12}$ \\
Virial Radius $R_{vir}$ (kpc) & 258 & 323 \\
Halo concentration $c$	& 12 & 9 \\
Disk Mass $M_{disk}$ ($M_\sun$) & $5.5\times10^{10}$ & $5.5\times10^{10}$ \\
Disk scale length $r_{disk}$ (kpc) & 3.5 & 3.5 \\
Disk scale height $z_{disk}$ (kpc) & $r_{disk}/5.0$ & $r_{disk}/5.0$ \\
Bulge Mass $M_{bulge}$ ($M_\sun$) & $10^{10}$ & $10^{10}$ \\
Solar distance $R_\odot$ (kpc) & 8.0 & 8.0 \\ 
Mass fraction of baryons in the disk & 0.055 & 0.0275 \\
\enddata
\label{Table:model}
\tablenotetext{a}{The Fiducial model is consistent with model A$_1$ of KZS02. }
\tablenotetext{b}{The High Mass model is consistent with model A$_4$ of KZS02 and is discussed in $\S$~\ref{subsec:mass}. }
\end{deluxetable*}

\begin{deluxetable*}{ccccc}
\tabletypesize{\scriptsize}
\tablecaption{Model Constraints}
\tablewidth{0pt}
\tablehead{
\colhead{Parameter} & \colhead{Observations} &  \colhead{Fiducial} & \colhead{High Mass} & \colhead{Other models}}  
\startdata  
Baryonic Mass inside $R_\odot$ ($10^{10}M_\sun$) & $>3.9$ [1] & 4.1 & 4.1  & 4.5 [2] \\
Escape Velocity at $R_\odot$ (km/s) & 550-720 [3] & 549 & 698 & 544 [4]\\
Total Mass inside 50 kpc ($10^{11}M_\sun$) & 3.2-5.5(4.0-6.4) [5]; & 3.8 & 5.0  & 3.7 [2]\\
  &  &  &  &  1.8-5.6 [6] \\   
Hot halo \# density at 50 kpc ($10^{-5}$ cm$^{-3}$) & $<50$ [7] & 25 & 80 & 1 [8] - 10 [9] \\
Total Mass inside 100 kpc ($10^{11}M_\sun$) & \nodata  & 6.0 & 9.1 &  6.0-6.6 [10]; 5.8 [2]\\
Total Mass within $R_{vir}$ ($10^{12}M_\sun)$ & 1.1-2.2(1.5-3.0) [11]; & 1.0 & 2.0 &  1.0 [2]; 0.88-2.56 [4] \\
 & 1.5 [12]; 0.2-5.5 [6] &  &  & \\
\enddata
\label{Table:Constraints}
\tablerefs{[1] \citet{BinEvans}: constraints from the number of microlensing events within $R_\odot$; [2] KZS02: model A$_1$; [3] \citet{Brown}: estimates from hypervelocity stars; [4]\citet{Smith}: quoted is their median likelihood; [5] \citet{Kochanek}: constraints from the local escape velocity of stars, disk rotation curve and the motions of satellite galaxies, excluding(including) Leo I; [6] \citet{Wilkinson}: motions of satellite galaxies and globular clusters at $r>20$ kpc; 
 [7] \citet{Rasmussen}: upper limit for the central halo density of the Local Group; [8] \citet{Murali}: lower limit from the lifetime of the MS;
 [9] \citet{Moore}: estimated from ram pressure arguments for the formation of the MS; [10] \citet{DehnBinn}: their models 1-4; [11] \citet{Sakamoto}: constraints from kinematics of MW satellites, globular clusters and horizontal-branch stars, excluding(including) Leo I;
 [12] \citet{Dehnen2006}: determined from radial velocity dispersion profiles of halo objects.}
\end{deluxetable*}

Note that if the MW were treated as a point mass, the escape velocity at any radius should be $\sqrt{2}$ times the circular velocity ($v_{circ}$). However, as emphasized by \citet{Smith}, the fact that estimates of the escape velocity at the solar radius are larger than $\sqrt{2}v_{circ}$ (see Table~\ref{Table:Constraints}) supports the existence of a massive dark matter halo. Thus it is also consistent that $v_{esc} \approx 380$ km/s $ > \sqrt{2}v_{circ} = 280$ km/s at 50 kpc: $v_{esc}$ probes the total mass rather than the enclosed mass.

We describe the gravitational drag induced by the perturbed density field resulting from the LMC's motion through the DM halo of the MW using the Chandrasekhar formula (\ref{eq:dynfric}):
\begin{equation}
\rm {\bf F}_{\rm DF} = - \frac{4\pi G^2M_{\rm sat}^2 \rm ln(\Lambda) \rho(r)}{v^2}\left[\rm {erf}(X) - \frac{2X}{\sqrt{\pi}}\rm {exp}(-X^2) \right] \frac{\bf {v}}{v},
\label{eq:dynfric}
\end{equation}
where $\rho(r)$ is the density of the host halo at the Galactocentric distance of a satellite of mass ${\rm M}_{\rm sat}$, $v$ is the orbital velocity of the satellite and X $= v/\sqrt{2}\sigma$. Here, $\sigma$ is the 1D velocity dispersion of the DM halo, which can be determined from the Jeans equation; we adopt the analytic approximation for an NFW profile derived by \citet{Zentner} (their equation 6).
Although the Chandrasekhar formula is strictly appropriate for the idealized scenario of an infinite, uniform, non self-gravitating stellar background, dynamical friction studies suggest that this approximation can be applied to more general situations \citep{Zentner, TB01}.
 Most previous studies of the orbital evolution of the Clouds have described the Coulomb logarithm as a constant value $\Lambda = b_{\rm max}/b_{\rm min}$ \citep{BT}, where $b_{\rm max}$ is a fixed cut-off radius and $b_{\rm min}$ is the impact parameter.
Instead, we adopt the prescription of \citet{HFM03}, wherein $b_{\rm max}$ is replaced by the radial position of the orbiting satellite at any given time and $b_{\rm min} = 1.4k$, where $k=3$ kpc is the softening length if the LMC were modeled using a Plummer profile.
 We show in the next section that, contrary to the results of all previous models, the influence of dynamical friction on the orbital evolution of the LMC is minimal for our fiducial model since the LMC has only recently entered the high density regions of the MW halo. Consequently, uncertainties in the analytic prescription for ${\rm F}_{\rm DF}$ do not affect our conclusions. Also, throughout this analysis we ignore the effects of tidal stripping since this process is similarly inefficient at large distances from the halo center. 

\section{Plausible Orbital Histories}
\label{sec:results}

In the following sections we present the results of our orbital analysis for our fiducial MW model and discuss possible constraints from the position of the MS on the plane of the sky. We further illustrate that our results are robust to changes in model parameters: namely, the mass and shape of the MW halo.  

\subsection{Discussion of Fiducial Model Orbits}
\label{subsec:fid}

Here we present plausible orbital histories for the LMC following the
methodology outlined in the previous section. In our analysis, 10,000
combinations of the west and north components of the proper motion
 ($\mu_W$,$\mu_N$) were randomly sampled within the error space of 
K1's measurements.
For each of these combinations, the 3D velocity was derived and the
orbit of the LMC was followed backwards in time for a Hubble time using 
our fiducial MW model. All
other parameter errors, e.g., that of the distance modulus and the
current position, are small in comparison to those of the proper
motions and do not affect this analysis.

\begin{figure}[t]
\begin{center}
\includegraphics[scale =0.43]{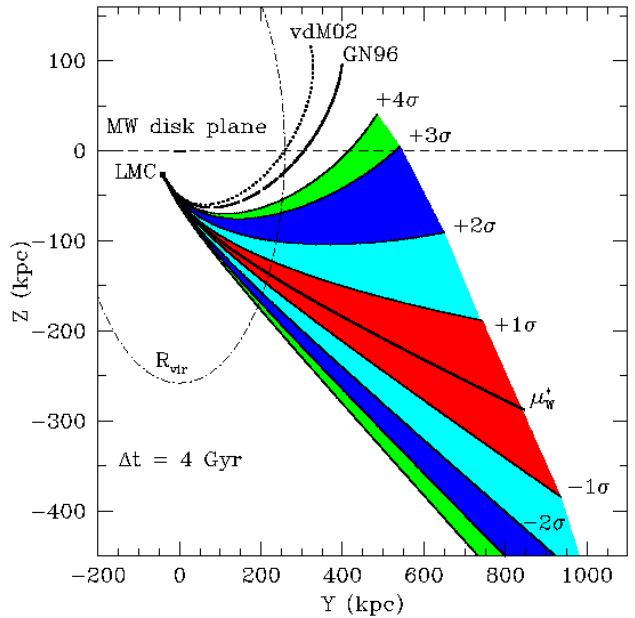}
\caption{Orbital paths corresponding to 10,000 randomly selected combinations ($\mu_W$,$\mu_N$) traced four Gyr in the past and plotted in the Galactocentric YZ plane. The present location of the LMC is indicated by the black square and the MW disk plane by the short-dashed line. The dash-dotted line indicates the virial radius of the halo (258 kpc). Orbital paths corresponding to $\mu_W$ values within $\pm1\sigma$ from the mean value ($\mu_W^* = -2.03$) are confined within the red area. Paths that correspond to $\mu_W$ components within (1-2)$\sigma$ are confined within the light blue region, between (2-3)$\sigma$ within the blue region and between (3-4)$\sigma$ within the green region.  Only orbits within the upper green region ($\mu_W = \mu_W^\ast+4\sigma$) cross the disk plane within 4 Gyr. The bold-face dashed and dotted lines indicate the orbits traced by the GN96 and vdM02 values, respectively, in our fiducial model.}
\label{fig:YZ}
\end{center}
\end{figure}

\begin{figure}[t]
\begin{center}
\includegraphics[scale=0.43]{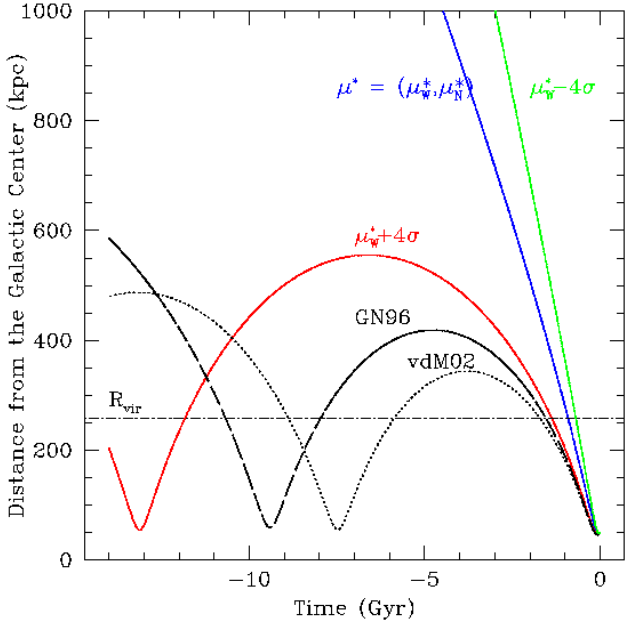}
\caption{Orbital evolution of the LMC plotted as a function of time in the past, where Time=0 corresponds to today. The dash-dotted line indicates the virial radius of the halo (258 kpc). All allowed orbits are constrained by the red and green lines, which indicate the outer boundary of the green regions in Figure~\ref{fig:YZ} ($\mu_W = \mu_W^\ast\pm4\sigma$). The blue line represents the orbital evolution of the LMC if the mean values are used ($\mu^\ast=(-2.03,0.44)$). Even in the best case scenario, the LMC completes only one orbit within $t_{\rm H}$ and reaches an apogalacticon distance of 550 kpc. The dashed and dotted lines indicate the orbits traced by the old GN96 and vdM02 values, respectively, in our fiducial model.}
\label{fig:OrbTime}
\end{center}
\end{figure}

In Figure~\ref{fig:YZ} the orbital path of the LMC is traced backwards
over four Gyr and plotted in the Galactocentric YZ plane. The
evolution of the galactocentric radius of the corresponding orbits are
plotted as a function of time in Figure~\ref{fig:OrbTime}, where $t=0$
corresponds to today. In both figures, the bold-face dashed(dotted)
line indicates the orbit traced by the old GN96(vdM02) velocities in
our fiducial model. Comparing these lines to their counterparts in
Figure~\ref{fig:Isothermal}, it is immediately apparent that the
orbital histories of the Clouds in a cosmologically-motivated MW model
are dramatically different from those in an isothermal sphere
model. In our fiducial model, the vdM02 and GN96 proper motions
(Table~\ref{Table:PMsummary}) imply that the LMC has completed only
one orbit within 10 Gyr and reached an apogalacticon distance of
300-400 kpc, whereas these same proper motions suggest orbital periods
of 1.5 Gyr and an apogalacticon distance of 100 kpc in the isothermal
sphere model. These striking results indicate that, independent of the
proper motion measurements, the choice of MW model can significantly
alter our picture of the orbital history of the Magellanic
system. The dependence of these results on model parameters are
discussed in $\S$\ref{subsec:mass}.

In Figure~\ref{fig:YZ} we also plot the orbital path of the LMC for each of the 10,000 proper motion combinations allowed within K1's error space. We find that the orbits are still roughly polar, but are not strictly confined within the YZ plane. 
Colored areas indicate the portion of the YZ plane spanned by orbits corresponding to ($\mu_W$,$\mu_N$) combinations within a specified $\sigma$ from the mean ($\mu^\ast$). The orbit corresponding to the mean values ($\mu^\ast$=(-2.03,0.44)) is parabolic. Orbits with a larger $|\mu_W|$ component (-1$\sigma$ to -4$\sigma$ from $\mu_W^\ast$) correspond to hyperbolic orbits. The magnitude of $|\mu_W^\ast|$ must decrease by 3-4$\sigma$ before the orbit will cross the disk plane. 
In Figure~\ref{fig:OrbTime} these same orbits are constrained between the red and green lines, which correspond to orbits with $\mu = \mu^\ast\pm4\sigma$. Even in the case most favorable to bound orbits 
($\mu_W = \mu_W^\ast+4\sigma$), the orbital period is roughly 
a Hubble time and reaches an apogalacticon distance of 550 kpc, which is substantially larger than that predicted by all other studies. On such distance scales dynamical friction plays little role in modifying the orbital history of the LMC and uncertainties in the analytic description of ${\rm F}_{\rm DF}$ are irrelevant ($\S$~\ref{sec:orbit}).

\begin{figure}[t]
\begin{center}
\includegraphics[scale=0.43]{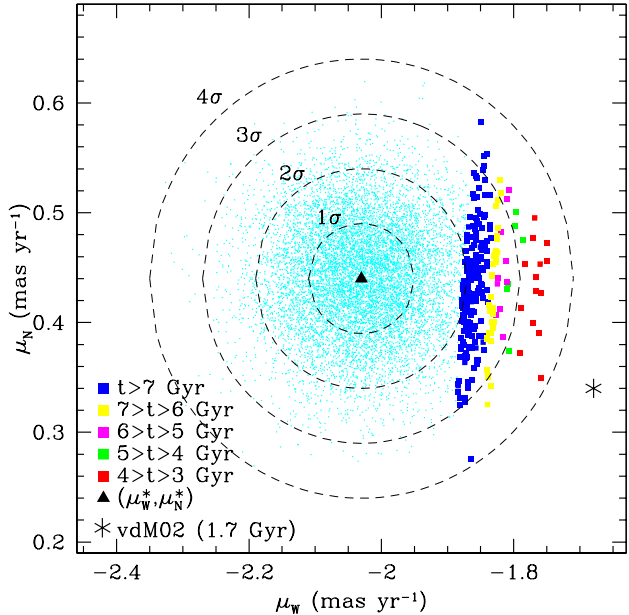}
\caption{10,000 points randomly sampled from the (4$\sigma$) proper motion error space of the K1 measurements for the LMC. The dashed ellipses indicate the standard deviation of the enclosed points from the mean (black triangle). For each point, the orbital history of the LMC was computed by integrating the equations of motion backward in time for the fiducial MW model. Solutions that cross the MW disk plane at least once within $t_{\rm H}$ are color coded by the time of the last disk plane crossing. Light blue dots indicate solutions that {\bf never} crossed the disk plane within $t_{\rm H}$. There are no solutions that cross $<$3 Gyr ago and in only $\sim$0.1\% of all cases does the LMC cross the disk plane $<4$ Gyr ago; i.e. twice the orbital period predicted by previous studies. The lack of colored points less than $\mu_W \sim-1.9$ indicates that the west component of the proper motion is the dominant factor in determining the orbital path of the LMC. The asterisk indicates the average of previous proper motion measurements as determined by vdM02 - the corresponding orbit crossed the disk plane 1.7 Gyr ago.}
\label{fig:PMTimeLMC}
\end{center}
\end{figure}

\begin{figure}[t]
\begin{center}
\includegraphics[scale=0.43]{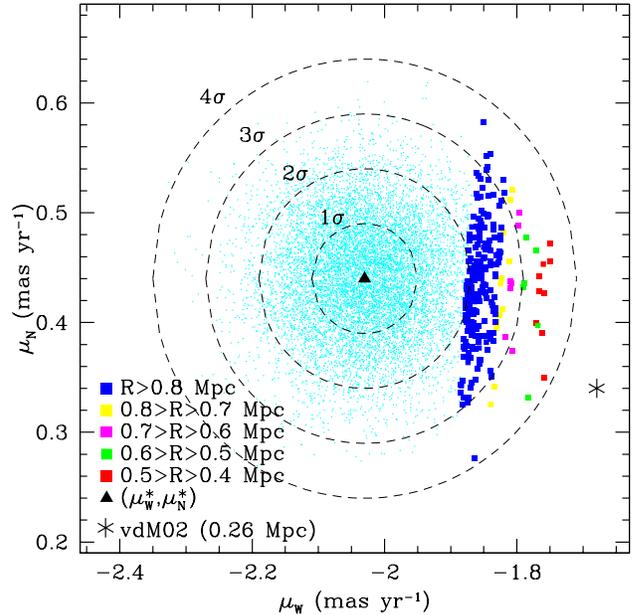}
\caption{Same as Figure~\ref{fig:PMTimeLMC}, except that solutions that cross the MW disk plane at least once within $t_{\rm H}$ (only 2\% of all cases) are color coded by the radial distance at which the LMC last crossed the disk plane.  Light blue dots again indicate solutions that {\bf never} crossed the disk plane. There are no solutions that cross the disk plane at distances smaller than 400 kpc; i.e., at four times the apogalacticon distance predicted by previous studies. The orbit corresponding to the vdM02 values crossed the disk plane at a distance of 260 kpc. }
\label{fig:PMDistLMC}
\end{center}
\end{figure}

We now examine these results more statistically: in Figures~\ref{fig:PMTimeLMC} and \ref{fig:PMDistLMC} all 10,000 combinations are color coded according to the time (Figure~\ref{fig:PMTimeLMC}) or distance (Figure~\ref{fig:PMDistLMC}) at which the LMC last crossed the MW disk plane. In both cases the light blue dots indicate orbits that {\bf never} crossed the MW disk plane within $t_{\rm H}$.  Dashed lines indicate the number of standard deviations of a given point from the mean proper motion (($\mu_W^\ast$,$\mu_N^\ast)$; black triangle). The large asterisk shows the vdM02 average of previous proper measurements: this value is well outside 4$\sigma$ of $\mu_W^\ast$, but is consistent with $\mu_N^\ast$. The distance and time of disk crossing for the vdM02 value (250 kpc; 1.5 Gyr ago) are marked in the legend.

No solutions within $1\sigma$ of the K1 mean ever cross the disk 
plane: in most of those cases the LMC never completes a single orbit
within $t_{H}$.  The 2\% of cases that do cross the disk plane do so
at substantially larger times and distances than predicted by any
previous study. In fact, only $\sim$0.1\% of all cases ever cross the
disk plane $<4$ Gyr ago; i.e. at times greater than twice the orbital
period predicted by GN96 using an isothermal sphere MW model. These
plots thus provide a measure for the timescale and strength of the
interaction between the MW and LMC.

From Figures~\ref{fig:PMTimeLMC} and \ref{fig:PMDistLMC}, the dominant factor controlling the orbital history of the LMC is the west component of the proper motion ($\mu_W$): if a solution exists for a given $\mu_W$ value, then it will exist for all $\mu_N$.
 The west component determines the tangential velocity - as the magnitude of $\mu_W$ increases, so does $v_{\tan}$ and, correspondingly, the orbital eccentricity. As such, the results for proper motion estimates with $\mu_N$ values beyond the scale of the y-axis can be extrapolated from the result of any proper motion combination with the same $\mu_W$.
For example, the GN96 proper motion estimate of ($\mu_W$,$\mu_N$) = (-1.72,0.12) is well outside 4$\sigma$ of K1's mean values and does not appear in our figures. But, by the above argument, we expect the GN96 results to fall somewhere between the $\mu_W+4\sigma$ results and those of the vdM02 proper motions. Reading from Figures~\ref{fig:PMTimeLMC} and \ref{fig:PMDistLMC}, if $\mu_W \approx -1.72$, the GN96 orbit should have crossed the disk plane $\sim1.7-3$ Gyr ago at a distance of $\sim 260-400$ kpc.
Checking Figures~\ref{fig:YZ} and \ref{fig:OrbTime}, we find that indeed the GN96 orbit crossed the disk plane at a distance of $\sim$300 kpc, $\sim$2 Gyr ago (the GN96 orbit spans only the YZ plane).   

\begin{figure}[t]
\begin{center}
\includegraphics[scale=0.43]{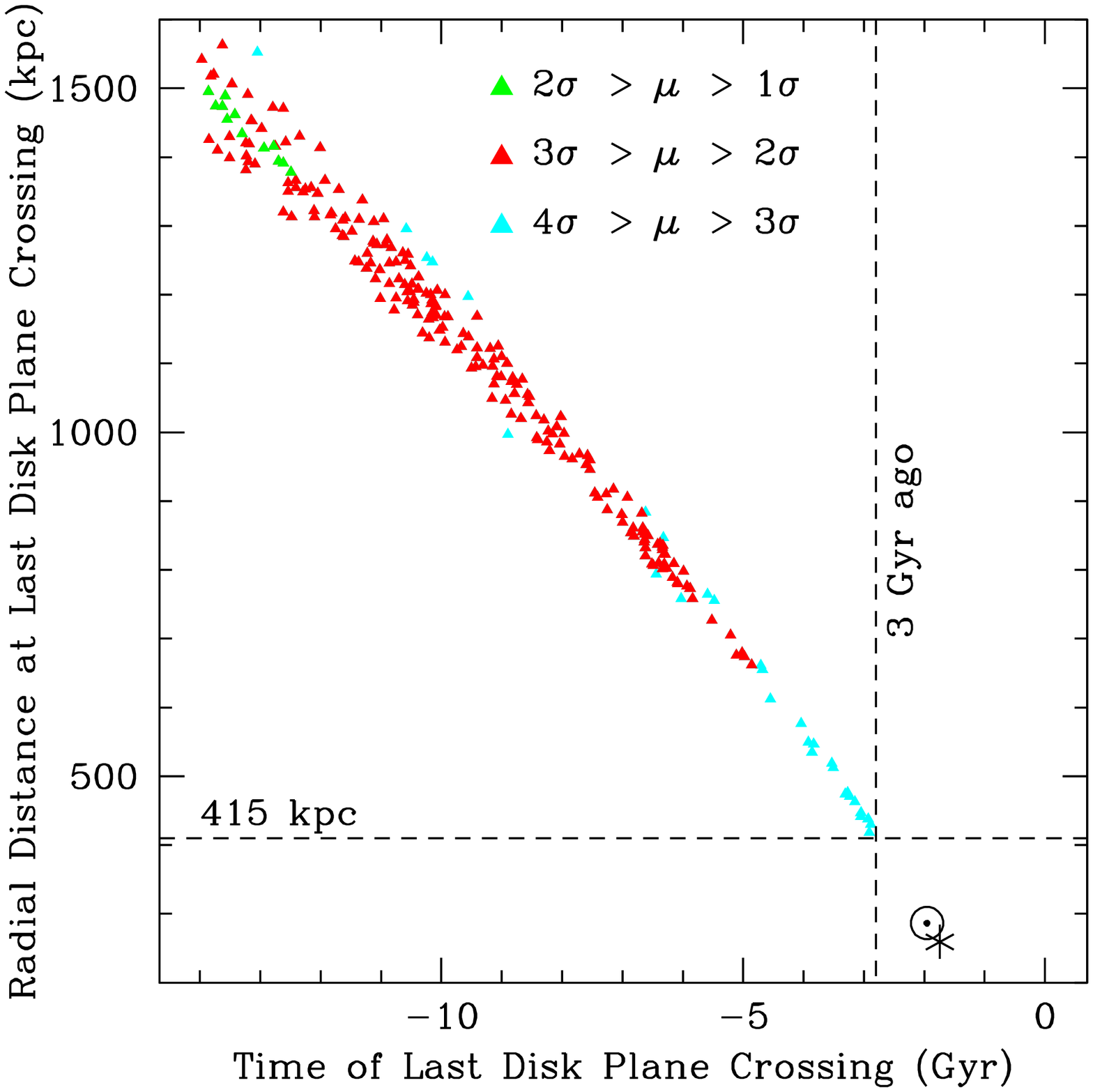}
\caption{This plot summarizes the results of Figures~\ref{fig:PMTimeLMC} and ~\ref{fig:PMDistLMC}. Here, all solutions that crossed the disk plane (the colored points from the previous two figures) are plotted as a function of the time and radial distance at crossing. Points are color coded by the deviation of the proper motion combination from the mean value: $\mu$ refers to some combination of $\mu_W$ and $\mu_N$, although in practice only $\mu_W$ matters. Only 0.1\% of cases within 1-2$\sigma$ of the mean ever cross the disk plane within $t_H$. Also, no orbits cross the disk plane $<3$ Gyr ago or at distances smaller than 415 kpc. The symbol $\odot$($\ast$) indicates the result for the GN96(vdM02) values in the fiducial model.}
\label{fig:TimeDistSigma}
\end{center}
\end{figure}

In Figure~\ref{fig:TimeDistSigma} the colored (disk crossing)
solutions from the previous figures are re-plotted as a function of
the time and radius of crossing. They are now color-coded by the
standard deviation of the proper motion from the mean value.  Even
within 4$\sigma$ of the mean, the LMC only crosses the disk plane at
timescales longer than three Gyr and at distances larger than 415 kpc.
Moreover, it is unreasonable to consider the LMC as an isolated system
or ignore cosmology over such distances and timescales.

We thus conclude that in our fiducial model the orbit of the LMC must
be close to parabolic: the LMC is on its first passage about the MW
and is currently at perigalacticon.

\subsection{Orbital Constraints from the MS}
\label{subsec:MS}

We now ask whether we can rule out either our fiducial (NFW) or
isothermal sphere MW models by comparing the projection of the
corresponding orbits of the LMC on the sky to the current position of
the MS. This exercise is motivated by the work of \citet{Johnston99},
who suggest that tidal streams act as ``fossil records'' of the recent
orbital history of their progenitors and could thereby provide a probe
of the galactic potential.  We note that the formation mechanism of the
MS is still a subject of debate; however, both the ram pressure and tidal 
stripping models predict that the orbit of the Clouds (approximated
here just by the LMC) should trace
the MS for at least some time in the past (e.g. GN96 and M05). 
Since the isothermal sphere and the fiducial (NFW)
model give very different orbital histories in 3D space, it is
worth exploring whether they give drastically different projected
locations for the LMC in comparison to the MS.

\begin{figure}[t]
\begin{center}
\includegraphics[scale=0.43]{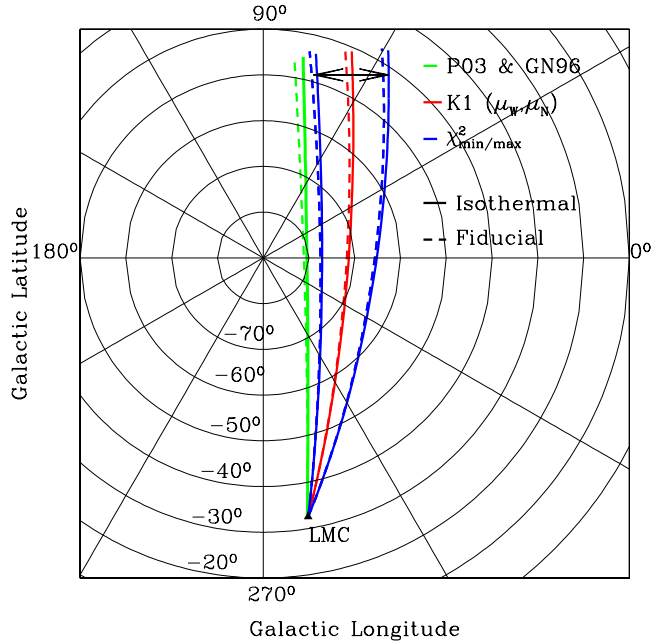}
\caption{Possible orbits of the LMC are mapped as a polar projection in galactic (l,b) coordinates. The orbit is followed backwards in time from the LMC's current position (black triangle) until it extends 100$\degr$ in the sky. Solid(dashed) lines indicate orbits computed using the isothermal sphere(fiducial) model. The green lines trace the orbit implied by the GN96 velocities: GN96 chose their initial velocities such that the LMC's orbit traces the the true position of the stream (P03). The red lines show the values using the mean K1 velocities, while the blue lines depict the best and worst fits of the allowed orbits to the position of the MS (max and min $\chi^2$, indicated by the black arrow).}
\label{fig:MS}
\end{center}
\end{figure}

  
Figure~\ref{fig:MS} shows the LMC orbits corresponding to the K1 and
GN96 velocities for both our fiducial and isothermal models as seen by
an observer situated at the location of the Sun looking towards the
South Galactic Pole. The orbits are traced backwards in time until the
LMC's line of sight position vector subtends an angle of $100\degr$ with
respect to its present-day position vector: this corresponds to the
extent of the MS \citep[hereafter {\bf P03}]{Putman2003}.

GN96 specifically chose 3D velocities for the LMC such that its orbit in an isothermal sphere MW model would trace the observed position of the MS (solid green line). We find little deviation in the projected orbital path of the LMC if our fiducial model is used with the GN96 velocities (dashed green line), even though the orbits differ substantially in 3D space (see Figures~\ref{fig:Isothermal} and \ref{fig:OrbTime}). 

The dashed(solid) red line traces the LMC's orbit using the mean velocities of K1 for the fiducial(isothermal sphere) model. Once again, in projection there is little deviation between the two MW models. There is, however, a striking difference between the allowed K1 orbits and the GN96 results: the LMC no longer traces the MS (solid green line).

The dashed(solid) blue lines represent the worst and best fits to the position of the MS for the fiducial(isothermal sphere) model. Orbits allowed within 4$\sigma$ of the mean are bounded by those lines, as indicated by the black arrow. These bounds were determined by extremizing the $\chi^2$ residuals of the LMC's orbit to a linear parametrization of the P03 HI data: Magellanic latitude B = 0.04L + 4.0.  The Magellanic coordinate system (L,B), defined such that its equator is parallel to the MS and the center of the LMC is currently located at L$\approx$-40 degrees, has been introduced here as a simple way to visualize the Magellanic system \citep{Wannier}.

 The good agreement between the projected orbits for our fiducial and
isothermal sphere models imply that we cannot use the location of the
MS to distinguish between MW models. Although a comparison of the
line-of-sight velocities of the orbit versus the HI data for the MS
could serve as a better discriminant, the correlation is strongly
dependent on the formation mechanism of the MS (see
$\S$\ref{subsubsec:stream}).

%

We have ignored the SMC so far in this analysis, even though previous models have assumed that the MS actually represents material torn from the SMC (e.g. GN96). The purpose of this section is two-fold: firstly, to show that we cannot differentiate between the NFW and isothermal halo models based on the current position of the MS; and secondly, to illustrate that the proper motion measurements necessitate a revision of current theories for the origin of the MS. The inclusion of the SMC, located 20 degrees from the LMC on the plane of the sky (or a separation of 23 kpc), does not affect either of these conclusions.

As we discuss in $\S$\ref{subsec:SMC}, the SMC's impact on the orbital history of the LMC is minimal, owing to the large LMC:SMC mass ratio of $\sim$10:1. Thus, the global dynamics of the LMC alone is sufficient to determine whether this analysis can differentiate between halo models (see also $\S$\ref{subsec:model}).  Furthermore, if we require that the Clouds have been bound to each other for at least some time in the past, the orbital path of the SMC will be strongly dependent on that of the LMC. Consequently, a shift in the LMC's orbit with respect to the current location of the MS will manifest as a shift in the orbital path of the entire Magellanic system. 
As such, the deviation cannot be explained away by assuming that the MS originated instead from the SMC (see Figure~\ref{fig:SMCpolar} in $\S$\ref{subsec:SMC}). 

\begin{figure}[t]
\begin{center}
\includegraphics[scale=0.8,angle=90]{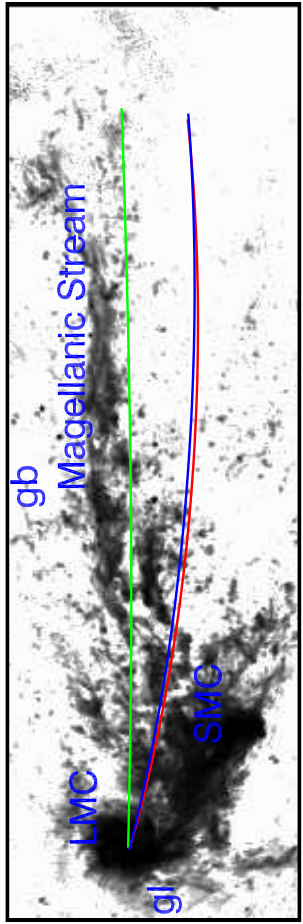}
\caption{The HI distribution of the MS from the data of P03 is plotted as a polar projection using the same scale as that in Figure~\ref{fig:MS}. The axis are the galactic longitude (gl) and latitude (gb). The SMC is located 20 degrees away from the LMC, corresponding to a physical separation of 23 kpc. The green line traces the orbit theoretically determined by GN96 in an isothermal MW model. Note that since GN96 advocate for a scenario in which the MS originates from the SMC, the green line does not match the data perfectly (see Figure~\ref{fig:SMCpolar} for a plot that includes the SMC's orbit). The blue(red) line shows the path traced by the vdM02(K1 mean) proper motions for the fiducial MW model. The vdM02 result is indistinguishable from that of the K1 mean values and both deviate by $\sim7\degr$ from the location of the MS: the observationally determined proper motions differ markedly from theoretical estimates.}
\label{fig:putman}
\end{center}
\end{figure}

We may also worry that this pronounced deviation implies an error in the K1 measurements. As mentioned in $\S$\ref{sec:intro}, HR94 consider a similarly large tangential velocity component as K1, yet their fiducial orbit traces the MS. This is because, like GSF94 and GN96, they {\it a priori} chose $\mu_N$ to ensure alignment. But these $\mu_N$ estimates are substantially different from not only the new K1 values, but also the weighted average of previous measurements as determined by vdM02 (see Table~\ref{Table:PMsummary}). Specifically, if we ignore the K1 values and repeat this analysis using the vdM02 proper motions instead, we get the {\it same result}: the vdM02 estimate of $\mu_N$ is similar to that of K1. Figure~\ref{fig:putman}, shows the projected orbital path of the LMC using the K1 and vdM02 proper motions (red and blue lines, respectively) overlaying P03's HI data for the MS. The K1 and vdM02 results are nearly indistinguishable, illustrating that the clear deviation in the projected orbital path and current location of the MS cannot be dismissed as an error in the new measurements.  

 As such, regardless of the choice of MW model, there are severe implications
 for either the tidal or ram pressure stripping theories for the origin of 
the MS:
the LMC's orbit using the K1 mean 
velocities deviates from the current location of the MS by $\sim$7
degrees on the sky, or $\sim$7 kpc if the MS traces a circle of radius
55 kpc. This deviation might be attributable to a number of second order
effects independent of the formation mechanism of the MS, e.g, 
the rotation of the MW halo or the L/SMC's disk might affect
the direction in which the material is removed. Regardless, such effects 
do not change our bottom line: all previous theoretical studies of the orbital
evolution of the Magellanic Clouds assumed that the MS and the LMC's orbit are co-located on the sky, meaning they are inconsistent with {\it both} the new {\it HST} and the weighted average of all previous proper motion measurements. The reliability of the K1 measurements will be discussed in detail in $\S$\ref{subsec:PM}.  

\subsection{Model Dependences}
\label{subsec:model}

Here we show that our analysis is robust to changes in model
parameters. Specifically, we examine whether the conclusions of
$\S$\ref{subsec:fid} change if we were to consider a MW model with
either a non-spherical or more massive halo than in our fiducial case.

\subsubsection{Mass Dependence}
\label{subsec:mass}

\begin{figure}[t]
\begin{center}
\includegraphics[scale=0.43]{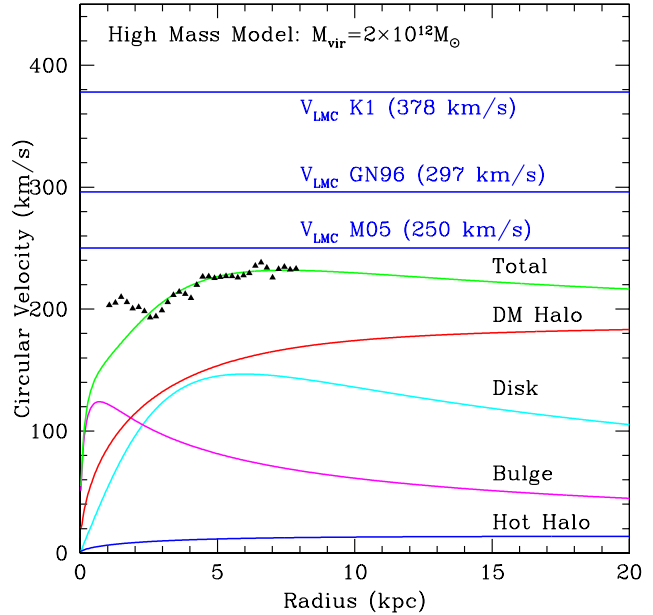}
\caption{Rotation curve plotted within 20 kpc for our High Mass MW model ($M_{vir}=2\times10^{10}\Msun$). The bulge and disk components are the same as in the fiducial model, while the halo mass has been increased by a factor of two. The NFW halo concentration parameter was decreased relative to the fiducial model in order to match the HI observational data within the solar radius (\citet{Knapp}; triangles). The escape velocity is now 552 km/s at 50 kpc, which is higher than the mean velocity estimate of K1 (top blue line). This model is consistent with KZS02's maximum halo mass model A$_4$.}
\label{fig:HighMass}
\end{center}
\end{figure}

We again consider a static 4-component MW model and maintain the same cosmological parameters. We also keep the same disk and bulge parameters, but increase the virial mass to $2\times10^{12}\Msun$. The halo concentration is decreased to $c=9$ in order to match the HI observational data of \citet{Knapp}. Model parameters are summarized in Table~\ref{Table:model} in the column labeled ``High Mass'' and the resulting rotation curve is plotted in Figure~\ref{fig:HighMass}. This model is consistent with KZS02's model A$_4$ ($M_{vir} = 2\times 10^{12}\Msun$, $c=10$). It also satisfies the observational constraints as listed in Table~\ref{Table:Constraints}, although the escape velocity at the solar radius is higher than the estimates of \citet{Smith} ($498<v_{esc}<608$ km/s) and the local number density of the gaseous halo conflicts with the upper limits of \citet{Rasmussen} (see Table~\ref{Table:Constraints}).  Note also that the escape velocity at 50 kpc is now 552 km/s, which is substantially higher than the mean velocity of K1.   

Higher mass models require even lower concentrations so that the maximal circular velocity does not exceed the observed value.  For example, if the upper limit of $M_{vir} = 5.5\times10^{12}\Msun$ \citep{Kochanek} is adopted, a concentration of $c=6-7$ is required. KZS02 rejected higher mass models based on the assumption that the concentration parameter should be in the range $c=10-17$; if $M_{vir}>2\times10^{12}\Msun$ and $c\ge10$, the maximum circular velocity would violate the HI observations. This range for the concentration parameter follows from $\Lambda$CDM model predictions that the virial mass and the concentration parameter are strongly correlated \citep{Bullock}. Moreover, for the \citet{Kochanek} upper mass limit, the escape velocity at the solar radius is $\sim1000$ km/s -- in comparison, the fastest hypervelocity star observed to date has a Galactic rest-frame velocity of 709 km/s \citep{Brown}.


\begin{figure}[t]
\begin{center}
\includegraphics[scale=0.43]{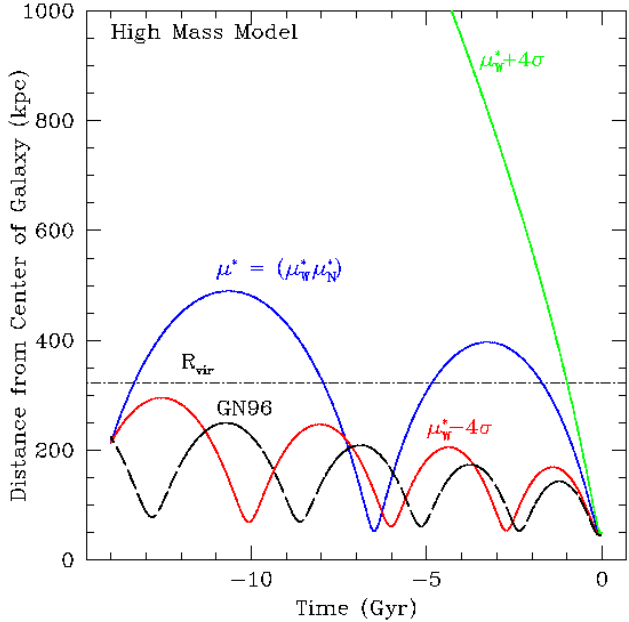}
\caption{Same as Figure~\ref{fig:OrbTime}, except that the orbital evolution is traced using the High Mass model. The dash-dotted line indicates the virial radius of the halo model (323 kpc). All allowed orbital histories are constrained by the red and green lines, which correspond to $\mu_W^\ast \pm 4\sigma$. The shortest allowed orbital period is $\sim3$ Gyr (red line). The orbit corresponding to the mean proper motions (blue) has an orbital period of $\sim$6 Gyr and reaches an apogalacticon distance of 400 kpc on the last passage. The orbit using the GN96 velocities in this model is also plotted for comparison (black dashed line); it is approaching the isothermal sphere solution of GN96 (see Figure~\ref{fig:Isothermal}). }
\label{fig:HighOrbitTime}
\end{center}
\end{figure}

We repeat the analysis of $\S$\ref{sec:results} using our High Mass model: the orbits of the LMC implied by the same 10,000 randomly selected $\mu_W$ and $\mu_N$ combinations are traced backwards for a Hubble time. The Galactocentric radial position of the LMC is plotted as a function of time in the past in Figure~\ref{fig:HighOrbitTime}, which complements Figure~\ref{fig:OrbTime} for the fiducial model. Although the results have improved, the orbital period for the mean values ($\mu^\ast$; blue line) is $\sim$6 Gyr and the LMC reaches an apogalacticon distance of $\sim$400 kpc. Thus, even in this High Mass model, the LMC travels on a highly eccentric orbit and is unlikely to have undergone more than one pericentric passage. In the best case ($\mu_W^\ast+4\sigma$; red line in Figure~\ref{fig:HighOrbitTime}) the LMC completes three orbits within the past 10 Gyr, but the  
apogalacticon distance is still $>150$ kpc. The orbit corresponding to the GN96 velocities in this High Mass model is shown for comparison (dashed black line).  Both these cases are approaching the isothermal sphere solutions of GN96, shown as the dashed red line in Figure~\ref{fig:Isothermal}. 

\begin{figure}[t]
\begin{center}
\includegraphics[scale=0.43]{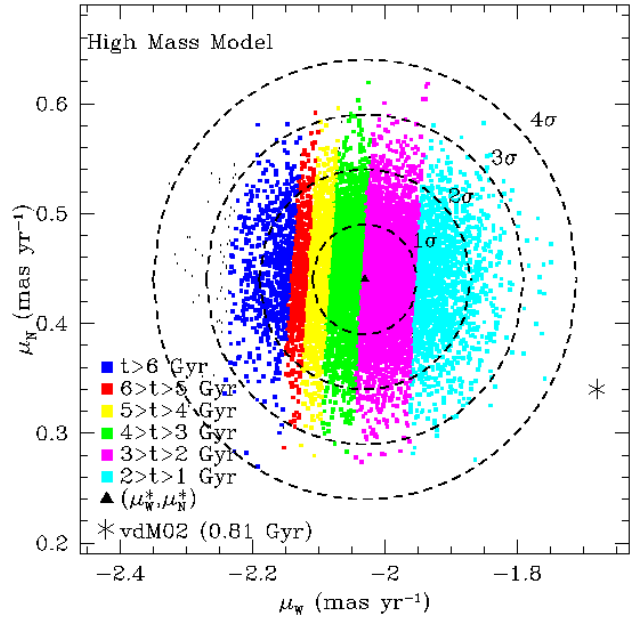}
\caption{Same as Figure~\ref{fig:PMTimeLMC}, except that orbits are computed for the High Mass model. Now 80\% of orbits cross the disk plane $<$4 Gyr ago: most orbits cross the disk plane at least once within $t_{\rm H}$, although if $\mu_W < (\mu_W^\ast - 3\sigma)$ solutions still cannot be found (black dots). Orbits within 1$\sigma$ do not cross the disk plane $<2.4$ Gyr ago and no orbits cross the disk $<$1 Gyr ago. Using the vdM02 values, the LMC crosses the disk plane 0.81 Gyr ago. }
\label{fig:HighMassTime}
\end{center}
\end{figure}

\begin{figure}[t]
\begin{center}
\includegraphics[scale=0.43]{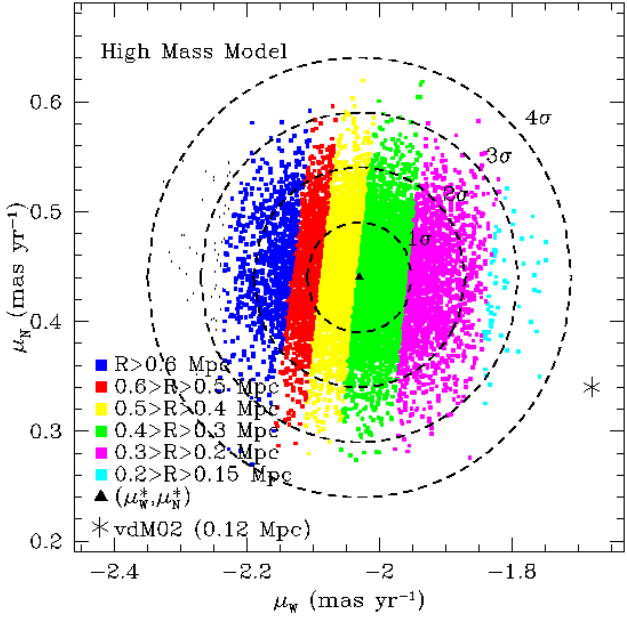}
\caption{Same as Figure~\ref{fig:PMTimeLMC}, except that orbits are computed for the High Mass model. Orbits within 1$\sigma$ do not cross the disk plane at Galactocentric distances $< 300 $ kpc and no orbits cross at radii $<150$ kpc. Using the vdM02 values, the LMC crosses the disk plane at a distance of 120 kpc.}
\label{fig:HighMassDist}
\end{center}
\end{figure}

We also repeat the statistical analysis of $\S$\ref{sec:results}: Figures~\ref{fig:HighMassTime} and \ref{fig:HighMassDist} complement Figures~\ref{fig:PMTimeLMC} and \ref{fig:PMDistLMC}, respectively. As before, orbits that cross the disk plane are color coded by the time or distance of the last crossing. As expected, there are more solutions that cross the disk plane at least once within $t_{\rm H}$ (80\% of cases cross the disk plane $<$4 Gyr ago). But, solutions still cannot be found for large $|\mu_W|$ components (black dots) and solutions within 1$\sigma$ of the mean do not cross the disk plane $<2.4$ Gyr ago or at distances $<$300 kpc, which corroborates our earlier results. 

We conclude that, if the LMC is moving with a velocity close to the mean value determined by K1 ($v=378$ km/s) and if the MW is not well approximated as an isothermal sphere at distances $\gtrsim$ 200 kpc, the LMC is unlikely to have undergone more than one pericentric passages about the MW within a Hubble time.  

\subsubsection{Axial Ratio of the MW halo}
\label{subsec:axial}

\begin{figure}[t]
\begin{center}
\includegraphics[scale=0.43]{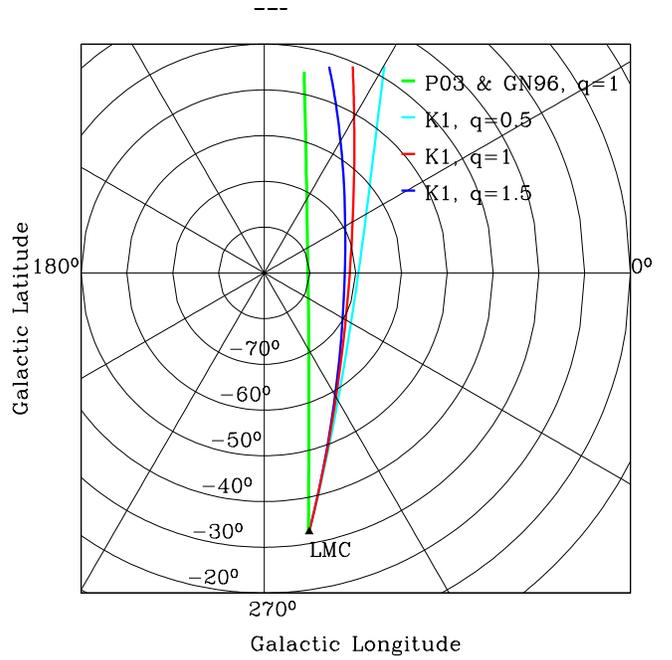}
\caption{Possible orbits of the LMC using only the mean K1 values mapped as a polar projection in galactic (l,b) coordinates for isothermal sphere MW models with varying axial ratios. The orbit is followed backwards in time from the LMC's current position (black triangle) until it extends 100$\degr$ in the sky. The green line traces the orbit implied by the GN96 velocities in an isothermal model of the MW. The other lines represent orbits in halos of varying sphericity: light blue (q=0.5; oblate), red (q=1.0; spherical) and blue (q=1.5; prolate).}
\label{fig:Axial}
\end{center}
\end{figure}

So far, we have considered only spherical halo models: if we vary the axial ratio, can a better fit between the position of the MS on the sky and our projected orbits be obtained? Given the good correlation between the projected orbits for our fiducial and isothermal sphere models, we assess the effects of halo sphericity using the isothermal sphere model of \citet{Binney81}:
\begin{equation}
\Psi = \frac{-v_o^2}{2}{\rm log}\left (R^2 + z^2/q^2 \right),
\label{eq:axial}
\end{equation}
where we have adopted cylindrical polar coordinates
($R$,$\phi$,$z$). The parameter $q$ defines the axial ratio of the
halo potential; in general, the corresponding density profile will be
more non-spherical. We vary $q$ between $q=0.5$ (oblate) and $q=1.5$
(prolate; spherical halos correspond to $q=1$)
 and compute the resulting orbits for the mean K1
 velocities\footnote[1]{We have checked the corresponding density
 profiles and ensured that they are reasonable. The oblate halos have
 the most non-physical density profiles and are slightly dimpled at
 the poles.}. The orbits are plotted in projection in
 Figure~\ref{fig:Axial}. We find that prolate halos provide better
 $\chi^2$ fits than oblate halos. However, the position of the orbit
 relative to the MS remains largely unaffected by the halo shape in
 the recent past - which is where the tidal and ram pressure stripping
 models predict the best agreement.

There are few constraints on the sphericity of the MW
halo. \citet{Fellhauer} determined that the angular differences on the
sky between the bifurcated stellar tidal streams of the Sagittarius
dwarf spheroidal detected by \citet{Belokurov} implied minimal
precession of the orbital plane. They concluded that the
axial ratio of the halo must be close to spherical. However, this
model does not explain the apparent velocity difference of $\sim15$
km/s between the bifurcated streams: \citet{Helmi} suggests that the
stream kinematics favor prolate halos, as determined from radial
velocities of M giants in the leading arm selected from 2MASS ({\it Two
Micron All Sky Survey}). On the other hand,
\citet{Johnston2MASS} used the 2MASS M giants to trace the tidal
streams and determined that the level of precession was best matched
by oblate halos.  Clearly there are no definitive
answers. Unfortunately, Figure~\ref{fig:Axial} shows that our analysis
cannot provide additional constraints on the halo sphericity. 


We go one step further and explore models in which, in addition to
varying the axial ratio, we also vary the slope, $\beta$, and
amplitude, $\eta$, of the rotation curve at $\sim 50$ kpc. We use as
our basis the ``power law'' galaxy model of \citet{Evans} in which the
potential is given by
\begin{equation}
\Psi = -\frac{\kappa}{(R^2 + z^2q^{-2})^{\beta/2}},
\end{equation}
where $\kappa = \frac{\eta^2 v_0^2}{\beta}[R_{\rm LMC}^2 + z_{\rm
    LMC}^2/q]^{\beta/2}$. This prescription gives a series of
    uninterrupted galaxy models with rising ($\beta < 0$) and falling
    ($\beta > 0$) rotation curves at large radii and properties
    abutting those of the isothermal sphere. We explore models with
    $-0.7< \beta < 0.7$ and $0.9v_0 < \eta < 1.1v_0$. However, after
    an extensive parameter search, neither of the parameters $\beta$
    or $\eta$ improve the agreement with the MS. The parameter $q$ is
    found to make the biggest difference to the shape of the orbit.

Since model parameters do not appear to be able to explain the
deviation between the projected position of the MS and the orbits
determined by the K1 velocities, a revision of current theories for
the formation of the MS may be warranted. The implications of our
analysis to formation mechanisms of the MS is continued in detail in
$\S$\ref{subsubsec:stream}.

\section{Discussion}
\label{sec:discuss}

In the following sections we show that our conclusions are robust to perturbations from the SMC ($\S$\ref{subsec:SMC}) and consider caveats to our analysis: namely the reliability of the recent proper motion measurements ($\S$\ref{subsec:PM}).  We further discuss the likelihood of a first passage scenario ($\S$\ref{subsec:first}) and the implications of the new velocity measurements for the star formation history of the LMC, the nature of the warp in the MW disk and the origin of the Magellanic stream ($\S$\ref{subsec:Imp}). Specifically, we illustrate that even if the MW were modeled as an isothermal sphere, there are significant challenges to our current understanding of the formation of the Magellanic stream (MS).

\subsection{The SMC-LMC Binary System}
\label{subsec:SMC}

The LMC:SMC mass ratio of $\sim$10:1 likely precludes the SMC from being a major determinant in the LMC's orbital evolution. However, observations of a common envelope of HI gas surround the L/SMC and a bridge of material linking the two Clouds imply that the L/SMC must have maintained a binary state for some time in the past. In this section we discuss the role of LMC-SMC interactions to the robustness of the predicted LMC orbits and the long-term stability of a binary LMC-SMC system. 

We again follow the methodology outlined in $\S$\ref{sec:orbit}, however the equations of motion for the LMC have been modified from equation~(\ref{eqmotion}) to include the gravitational interactions between the SMC and LMC. We model the LMC(SMC) using a Plummer profile with a softening parameter of 3(2) kpc. The SMC mass is assumed to be $2\times10^{9}\Msun$ (SMC:LMC mass ratio of 1:10). The equation of motion for the SMC is analogous, except that we also include a dynamical friction term that acts on the SMC when it enters the current tidal radius of the LMC ($\sim$15 kpc; \citet{vanderMarel}). We follow the analysis of \citet{Bekki} and approximate this friction term as \citep{BT}:
\begin{equation}
\rm {\bf {F}}_{\rm LS} = -0.428 \ln(\Lambda_{\rm L,S}) \frac{G\rm{M}_{\rm S}^2}
{r_{\rm LS}^2}\frac{\bf{v}_{\rm LS}}{v_{\rm LS}},
\label{DynFricBT}
\end{equation}
where $v_{\rm LS}$ is the relative velocity between the Clouds and the Coulomb logarithm is $\rm ln(\Lambda_{\rm L,S}) =0.2$. 

\begin{figure}[t]
\begin{center}
\includegraphics[scale=0.43]{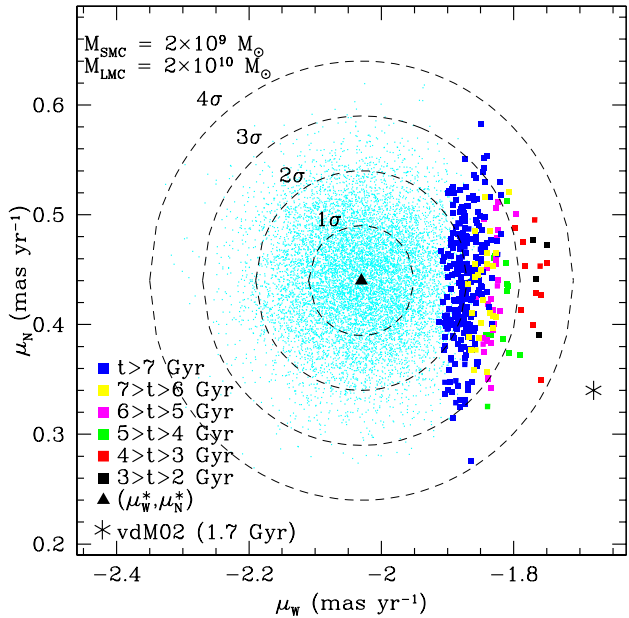}
\caption{Same as Figure~\ref{fig:PMTimeLMC} except that interactions between the LMC and SMC have been included. The SMC introduces scatter, but the previous results are unchanged: there are no solutions that cross the MW disk plane within $1\sigma$ of the mean or for large $|\mu_W|$.}
\label{fig:SMCPMTime}
\end{center}
\end{figure}

\begin{figure}[t]
\begin{center}
\includegraphics[scale=0.43]{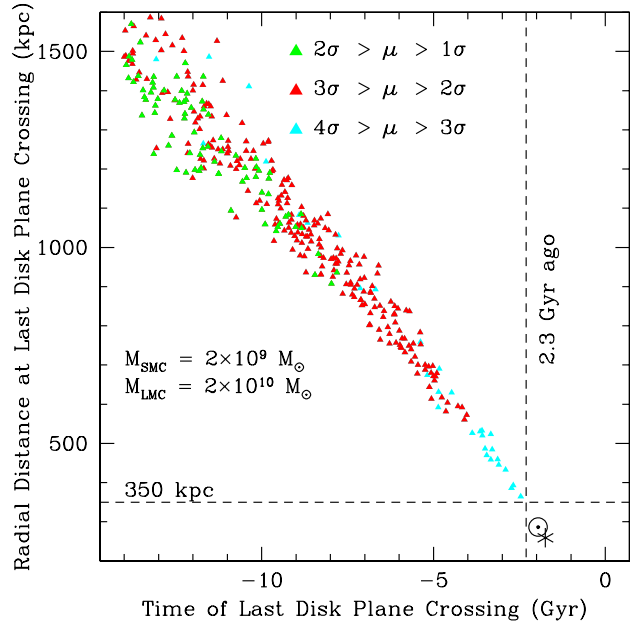}
\caption{Same as Figure~\ref{fig:TimeDistSigma} but including interactions between the LMC and SMC. The plot is more scattered but still consistent with the previous results: there are no cases that crossed the disk plane $<$2.3 Gyr ago or at distances $<$350 kpc and the best results are obtained if $\mu > (\mu^\ast + 3 \sigma)$. The symbol $\odot$($\ast$) indicates the result for the GN96(vdM02) values.}
\label{fig:SMCPMTD}
\end{center}
\end{figure}

Here, we repeat the statistical analysis of $\S$\ref{sec:results} with
the inclusion of the SMC. 10,000 pairs of points were drawn randomly
from both the LMC and SMC proper motion error space (K1,K2). The
corresponding SMC and LMC orbits were traced backwards in time for a
Hubble time in our fiducial MW model. In Figures~\ref{fig:SMCPMTime}
and \ref{fig:SMCPMTD} we re-plot Figures~\ref{fig:PMTimeLMC} and
\ref{fig:TimeDistSigma} now including the SMC. The SMC introduces more
scatter into the plots and there are now more solutions within
2$\sigma$, but they exist at large distances and over long
timescales. Thus, the overall results are not substantially changed:
there are still no orbits within 1$\sigma$ that cross the disk plane.

Although there are large uncertainties in the LMC:SMC mass ratio, increasing the SMC mass by a factor of a few does not substantially change these results. The same is true if this analysis were repeated using the High Mass model described in $\S$\ref{subsec:mass}.

The calculations of \citet{Bekki} for an isothermal sphere MW model suggest that it is unlikely that the LMC and SMC remained a bound system for a Hubble time since dynamical friction owing to the SMC's passage within the LMC's halo destabilizes the binary system. In support, \citet{Bekki2} argue that the different cluster formation histories of the LMC and SMC may be explained if they originated as separate entities (see also $\S$\ref{subsec:Imp}). However, given the sparse distribution of the outer MW satellites, a capture scenario of the SMC by the LMC seems improbable.  We now examine whether the LMC and SMC can maintain a stable binary state in our fiducial MW model if the LMC is moving at the K1 mean velocity of 378 km/s.  

\begin{figure}[t]
\begin{center}
\includegraphics[scale=0.43]{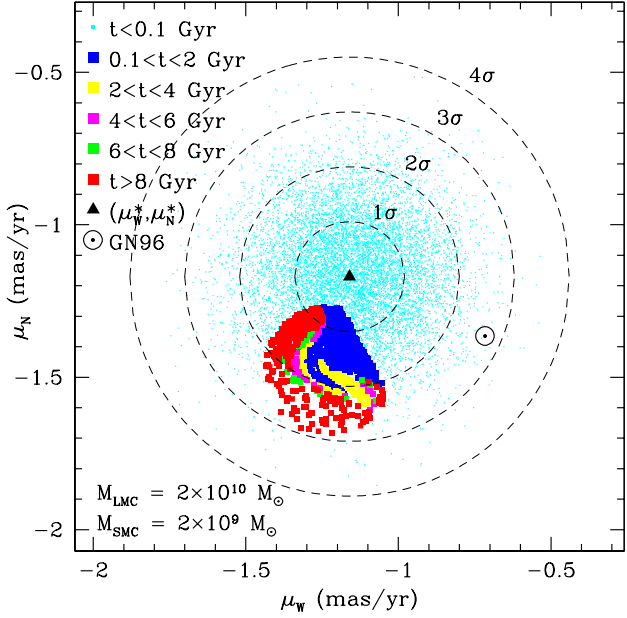}
\caption{10,000 points randomly drawn from the SMC proper motion error ellipse (K2). Each point is color coded by the amount of time the SMC remains bound to the LMC, i.e. how long  $v_{esc} > v_{\rm LS}$, if the LMC is moving at the K1 mean velocity. In the majority of cases, this criterion is not currently satisfied (light blue dots). The red squares indicate solutions where the binary-system is maintained for $t_{\rm H}>t>8$ Gyr: such solutions do exist within 1$\sigma$ of the mean (black triangle). The symbol $\odot$ indicates the proper motion corresponding to the GN96 estimate of the SMC's current velocity: in our model, this value corresponds to a currently unbound LMC-SMC system. }
\label{fig:Bound}
\end{center}
\end{figure}

10,000 points were randomly drawn from the SMC proper motion error
circle (K2) and plotted in Figure~\ref{fig:Bound}: they are color
coded based on the longevity of the binary state. Bound orbits are
identified by whether the SMC's velocity relative to the LMC ($v_{\rm
LS}$) is smaller than the local escape velocity. In most cases this
criterion is not initially satisfied (light blue dots; t$<$0.1 Gyr);
i.e. the SMC is currently not bound to the LMC.  But solutions where
the binary state is maintained for $8<t<t_{\rm H}$ (red squares) are
allowed within 1$\sigma$ of the mean.  Thus, although dynamical
friction between the clouds was included, binary states that last for
a Hubble time are plausible, which is contrary to the results of
\citet{Bekki}. This is because the relative velocity between the LMC
and SMC is determined by the tidal force imparted by the MW, and our
fiducial model differs significantly from the isothermal sphere model
used by \citet{Bekki}. Correspondingly, very few stable binary states
exist when this analysis is repeated for our High Mass model and it
becomes easier to maintain the binary state if the LMC:SMC mass ratio
is increased. This is consistent with a similar analysis by K2 in an
isothermal sphere model (i.e. a high mass model), in which they found
$<$ 10\% of orbits to be in a stable binary state.

\begin{figure}[t]
\begin{center}
\includegraphics[scale=0.43]{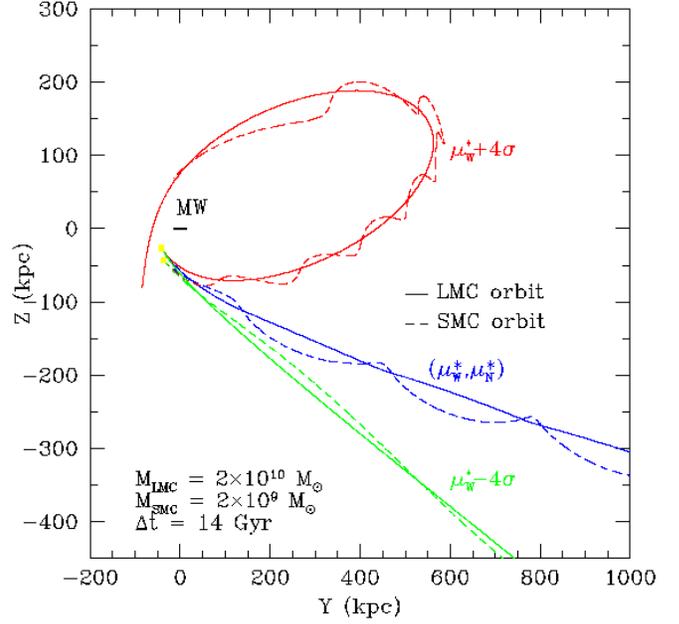}
\caption{Stable binary orbits plotted in the galactocentric YZ plane for the mean (blue) and limiting LMC velocities, which correspond to $\mu_W = \mu_W^\ast\pm4\sigma$ (red and green). The orbits are traced over a Hubble time. The LMC orbit is indicated by the solid line and the SMC by the dashed line. The orbital path of the LMC is not significantly affected by the SMC and a stable binary state can be found within the SMC proper motion error space for all allowed LMC velocities.}
\label{fig:BoundOrbit}
\end{center}
\end{figure}

Repeating this analysis using the limiting 4$\sigma$ LMC velocities (corresponding to $\mu_W = \mu_W^\ast\pm4\sigma$) instead of the mean LMC velocity, we find that a stable binary state ($t_{bound}>8$ Gyr) can always be located. However, it becomes increasingly difficult to identify stable binary states as the $|\mu_W|$ component is increased. The requirement of binarity could thus be used to further constrain the SMC or LMC proper motion error space: in order for the L/SMC to form a binary system their transverse motions must be comparable, and the new transverse motions for the LMC have increased substantially. 
In Figure~\ref{fig:BoundOrbit} the orbital paths of a stable binary system for the mean and limiting LMC velocities are traced over a Hubble time in the Galactocentric YZ plane.  We conclude that for every LMC orbit that is allowed by the data, there also exists a bound SMC orbit that is also allowed by the data.

\begin{figure}[t]
\begin{center}
\includegraphics[scale=0.43]{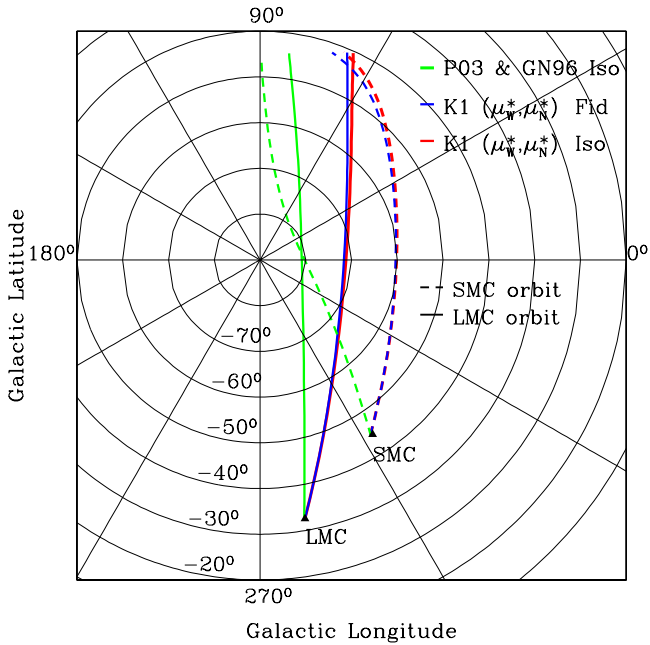}
\caption{ Possible orbits of the LMC (solid lines) and SMC (dashed lines) are mapped as a polar projection in galactic (l,b) coordinates. The orbits are followed backwards in time from the Clouds' current positions (black triangles) until they extend 100$\degr$ in the sky. The green lines trace the orbits implied by the GN96 velocities in an isothermal model of the MW (Iso). The red and blue lines show the projected orbits for the most long-lived binary state if the LMC is moving at the mean K1 velocities in the isothermal and fiducial MW models, respectively. }
\label{fig:SMCpolar}
\end{center}
\end{figure}

Considering only long-lived binary orbits (red squares in Figure~\ref{fig:Bound}), we now repeat the analysis of $\S$\ref{subsec:MS} and confirm the statement that the inclusion of the SMC cannot explain the discussed deviation in the orbital path of the Magellanic system with respect to the current location of the MS.
 In Figure~\ref{fig:SMCpolar} the projected path followed by the SMC (dashed line) and LMC (solid line) using the old GN96 values in an isothermal sphere model (green) are plotted in Galactic coordinates. Notice that the SMC's path traces the true location of the MS (shown in Figure~\ref{fig:putman}) better than the LMC does. However, for the allowed K1 LMC velocities, the GN96 estimate of the SMC proper motion (indicated by the symbol $\odot$) does not yield a stable binary LMC-SMC system. Instead, we plot the most stable binary state if the LMC is moving at the mean K1 values in the isothermal sphere (red) and fiducial (blue) MW models. Here the fiducial model orbits also correspond to the orbits traced by the blue lines in Figure~\ref{fig:Bound}. As expected, the deviation of the LMC's orbit translates into a shift in the orbital path of the entire Magellanic system. Thus, if the LMC and SMC formed a binary system for some time in the past, the LMC's orbit provides a fair assessment of the behavior of the entire Magellanic system.

\subsection{Reliability of the New Proper Motions}
\label{subsec:PM}




There are three new ingredients in this analysis: K1's $\mu_W$,
$\mu_N$ and the MW mass model. The far fewer number of perigalactic
passages for the Clouds is a generic prediction of our fiducial model
for the MW and is essentially independent of the proper motion
data. However the conclusion that the Clouds are on their
\textit{first} passage relies on the K1 $\mu_W$ value and
corresponding error bar.
 Given the unexpected implications for
the history of the MW-LMC-SMC system (see $\S$\ref{subsec:Imp}), it is
worth re-examining the new $\mu_W$ value in the context of other
work. K1's value for $\mu_W$ of the LMC is on the high end of the
values reported previously (see Figure~14 in K1), but
it is consistent within the errors with the $Hipparcos$ value
\citep{Kroupa97} and the \citep{Pedreros02} value; see also
Table~\ref{Table:PMsummary}.

The line-of-sight velocities of LMC carbon stars can also be used
to put constraints on the transverse velocity of the LMC center of
mass. Even though the rotation curve and the
transverse velocity component are not uniquely determined by the
line-of-sight velocity field, plausibility arguments based on the
Tully-Fisher relation and comparison to M33 puts the transverse
component between 200 and 600 km/s (vdM02). Thus the transverse velocities
used in this analysis are not ruled out by the carbon star
kinematics. An increase in the east motion of the LMC gives \textit{better
agreement} with the line-of-sight velocity field than the vdM02 proper motions,
as can be seen from
Figure~8 in vdM02. The dotted trapezoid in this figure
is the 68.3\% confidence region obtained from the line-of-sight
velocity field under the assumption that the rate of change of
inclination of the LMC disk, $di/dt$, is zero. In K1 a
combined analysis of the carbon star kinematics and the proper motion
field is used to show formally that $di/dt$ is more consistent with
zero than with the use of previous (lower) proper motion values.

In a recent analysis comparing the kinematics of HI, carbon stars and
red supergiants in the LMC (\citet{Olsen}), the authors were able to
correct the line-of-sight velocities for the K1 proper motion and make
a useful interpretation of the resulting velocity fields, finding that
each of these tracers defines a flat rotation curve. The large
tangential component of the LMC velocity does not introduce spurious
motions in the analysis of their line-of-sight data. Finally, a
substantial tangential component for the LMC was predicted by
\citet{Lin}, in which they explore a tidal model for the origin of the
MS in a massive dark halo model for the MW. Tangential velocities
consistent with these values were also predicted by HR94 in a ram
pressure stripping model for the MS (also with a massive MW halo).

Notably, K1's estimate of $\mu_N$ \textit{is} consistent with that of vdM02
(see Table~\ref{Table:PMsummary}). This implies that the apparent
shift in the orbital path of the Magellanic system with respect to the
current location of the MS, as discussed in $\S$\ref{subsec:MS},
exists {\it even if the old proper motion measurements are used} (see
Figure~\ref{fig:putman}). Theoretical models have invoked smaller
values of $\mu_N$ in order to satisfy the underlying assumption that
the Clouds should trace the MS.  However, this assumption is strongly
dependent on the formation mechanism of the MS, which in itself is
highly uncertain (see $\S$\ref{subsubsec:stream}).


\subsection{Evidence for a First Passage Scenario}
\label{subsec:first}

The possibility that the LMC is on its first passage about the MW is not a novel idea. As noted by \citet{Fich},\citet{Lin95} and \citet{Sawa}, if the LMC is currently at perigalacticon, its orbital angular momentum is comparable to the rotational angular momentum of the MW's stellar disk. 
 Since the LMC's orbit is near polar, the spin axis of the MW disk is perpendicular to that of the LMC's orbit. Thus, if the LMC has been in a decaying quasi-periodic orbit about the MW for a Hubble time, the origin of its high angular momentum is puzzling. The new higher tangential velocity measured by K1 only exacerbates this ``angular momentum problem'' and cannot be explained by current models of the orbital evolution of the Clouds. 

\citet{Raychaud} suggest that tidal torques exerted by M31 early in the LMC's orbital history may be responsible for the LMC's high tangential velocity. Building on this theory, \citet{Shuter} and \citet{Byrd} considered the possibility that the Clouds underwent a close encounter with M31 and were only recently tidally captured by the MW.  Although our computed orbits do not support a scenario where the LMC originates from M31,
 the source of the LMC's high tangential velocity may be easier to understand if the LMC was not always bound to the MW. In this way, the evolution of the LMC-SMC-MW system cannot be considered in isolation, and the LMC-SMC binary may have been subjected to external torques before entering the MW's virial radius.  

In addition, based on a morphological comparison of the satellite galaxies of the MW and M31, \citet{vandenbergh2006} suggests that the L/SMC may be interlopers that originally formed in a more remote region of the Local Group. All inner satellites of M31 are early-type objects, which is also true of the MW except for the presence of the irregular LMC and SMC at small galactocentric radii with respect to the other MW satellites. Moreover, as noted by \citet{Fich}, the maximum line of sight velocity of the MS ($v_{\rm LSR}$ = -410 km/s; \citet{Bruens2005}) is not only higher than that of the Clouds themselves (262 $\pm$ 3.4 km/s; vdM02) but also of any satellite within 200 kpc.  Thus, if the orbital path and velocities of the LMC are similar to that of the MS, they conclude that its orbits must be very different from those of other satellites (see also the discussion in $\S$~\ref{subsubsec:stream}). 

Following the analysis of \citet{Lin95} we next consider simple consistency checks for a scenario where the LMC is currently unbound to the MW. If the LMC's orbital eccentricity is negligible, the dynamical mass of the MW within the orbit of the LMC can be approximated as follows:
\begin{equation}
M_{\rm MW} \approx \frac{v_{tan}^2 R_{\rm{LMC}}}{G} = 1.6\times 10^{12} \Msun,
\label{eq:dynmass}
\end{equation}
where $v_{tan}$ is the tangential velocity of the LMC (367 km/s; K1) and $R_{\rm LMC}=50$ kpc is the current Galactocentric distance of the LMC. This is much larger than observational constraints (see Table~\ref{Table:Constraints}) and so the LMC cannot be on a circular orbit. 

Since the radial velocity is non-zero ($v_{rad}=89$ km/s; K1), we know the LMC must be on an eccentric orbit. Assuming that all the galactic mass is within 50 kpc, a lower bound on the dynamical mass can be estimated as:
\begin{equation}
M_{\rm MW} = \frac{R_{\rm LMC }}{2G}\left[ \frac{v_{\rm rad}^2 + v_{\rm tan}^2(1- R^2_{\rm LMC}/r_a^2)}{(1-R_{\rm LMC}/r_a)} \right ] = 8.4\times10^{11} \Msun
\label{eq:dynmassecc}
\end{equation}
where $r_a$ is the apogalacticon distance.  Following the analysis of \citet{Lin95}, if the LMC is bound to the MW then the apogalacticon distance must be less than the Galactic tidal radius with respect to M31, so $r_a<$300 kpc. The corresponding mass estimate is roughly the upper limit on the mass estimates within 50 kpc ($6.6\times10^{11}\Msun$ within 50 kpc; \citet{Wilkinson}). This simple lower bound suggests that the LMC is likely on a highly eccentric orbit and could be effectively unbound.

A first passage scenario for an LMC-type galaxy at the present epoch is not at odds with current theories of the hierarchical build-up of dark matter (DM) halos. In a study using both N-body simulations and Extended Press-Schechter calculations to estimate the fraction of MW size halos that have experienced recent mergers, it is found that 70\% have accreted an LMC-sized object in the past $\sim$5 Gyr (James Bullock, private communication 2007). Here ``accreted'' means ``first falls within'' a $\sim$300 kpc virial radius. For our fiducial model, the LMC typically enters the virial radius within the past 1-2 Gyr. The corresponding fraction of such mergers in the simulations is 30\%.
 We thus conclude that a first passage scenario for an LMC-type object is not statistically improbable. 


Finally, it should be noted that the primary constraint on the long-term orbital history of the Clouds comes from current formation mechanisms of the MS, which require that the Clouds have undergone multiple pericentric passages, and from theoretical links between the observed star formation rates and close encounters between the LMC-SMC-MW system. These issues are explored in detail in the next section.

\subsection{Phenomenological Implications}
\label{subsec:Imp}

In the following we consider the implications of the new proper motion measurements and consequent orbital history of the Clouds for our understanding of their star formation histories, the formation of the warp in the HI layer of the MW disk, and the origin of the MS.

\subsubsection{Star Formation History of the Magellanic Clouds}
\label{subsubsec:star} 

The star formation history (SFH) of the Magellanic Clouds is believed to be strongly influenced by dynamical and hydrodynamical galaxy interactions \citep{Westerlund}.  Detailed knowledge of the orbital history of the Clouds is thus critical to the understanding of their SFHs. In the picture we have developed in $\S$~\ref{sec:results} (our fiducial MW model), the Magellanic clouds are on their first passage about the MW and entered within the virial radius of the MW DM halo only $\sim(1-3)$ Gyr ago.  As they travel supersonically (mach $\sim$3) through the halo gas, they interact hydrodynamically with the ambient medium, forming a bow shock that will increase the gas density at the leading edge \citep{DeBoer}. As the Clouds approach the galactic center they are also subject to tidal forces from the MW. We therefore expect the Clouds to exhibit a heightened SFR within the past few Gyr. 

The SFH of the LMC's disk is believed to be relatively smooth and continuous, whereas the recent SFH of its bar is more episodic. Specifically, \citet{Smecker} identified the dominant stellar populations in the bar with episodes of star formation that occurred 4-6 and 1-2 Gyr ago. These episodes have been attributed to tidal interactions during pericentric passages between the LMC, SMC and the MW.
Moreover, an ``age gap'' has been observed in the LMC's globular clusters, which refers to the identification of only one cluster in the LMC with an age between $\sim$3-13 Gyr \citep{Rich, DaCosta}. \citet{Bekki} note that the reactivation of cluster formation $\sim$3 Gyr ago could be explained by the onset of hydrodynamic interactions between the LMC's disk and the ambient halo gas. The resulting bow-shock could act as a catalyst for cluster formation (see also \citet{DeBoer}). If the Clouds have undergone multiple pericentric passages about the MW, it is unclear why such hydrodynamical effects would be important only recently -- whereas this picture is a natural consequence of a first passage scenario.  


A rise in the SMC's mean SFR within the past 3 Gyr has also been inferred by \citet{Harris04}, who describe the recent star formation rate as highly irregular with bursts of star formation occurring at 2.5, 0.4, and 0.06 Gyr ago.
Unlike the LMC, the SMC is unbarred and has no spiral structure \citep{Zaritsky2000}, thus the mechanism for driving star formation in the SMC likely involves galaxy interactions. Specifically, \citet{Zaritsky04} suggest that the irregular appearance of the SMC principally owes to recent star formation triggered by hydrodynamic interactions between the SMC's gaseous components and the halo gas (or outer gas envelope of the LMC), rather than tidal effects. Thus the recent SFH of the SMC also appears to support a first passage scenario.



\subsubsection{Warp of the MW Disk}
\label{subsubsec:warp}

From 21-cm observations it is clear that the HI layer in the MW disk is substantially warped (vertically distorted to distances $>$3 kpc above the galactic plane; \citet{Diplas}). The dust layer has also been observed to be similarly displaced from the galactic plane \citep{Freud}.
Although the origin of the warp is still unknown, the LMC has historically been considered as providing a probable excitation mechanism owing to its proximity and mass.  However, even if the LMC were on a circular orbit of radius 50 kpc, its tidal fields are not strong enough to distort the HI MW disk by $>2$ kpc vertically \citep{Burke,Kerr,Hunter}. This could be improved if the LMC were substantially more massive than previously estimated ($>10^{11}\Msun$; \citet{Binney92}) or if the pericenter of its orbit were much smaller ($\sim$20 kpc; \citet{Hunter}). Our analysis suggests that the LMC is on a parabolic orbit and is only {\it now} at its closest approach to the MW: on such an orbit, the torque exerted by the LMC is insufficient to excite a warp of the observed magnitude. Thus, if the Magellanic Clouds are the source of the warp, their influence on the MW needs to be amplified. 

\citet{Weinberg98} proposed that the warp results from a combination of two processes: 1) direct tidal forcing by the LMC; 2) resonant forcing of disk bending modes by the density disturbance excited by the LMC's passage through the halo. 
If the LMC follows a non-decaying quasi-periodic orbit, as assumed by \citet{Weinberg98} and recently by \citet{Weinberg06},
 it will excite discrete modes in the halo. The higher order modes decay quickly, whereas low order modes persist over a few dynamical times;
 moreover, on a quasi-periodic orbit the weakly damped modes will be continually re-excited and so dominate the response. 
\citet{Tsuchiya} tested this scenario by means of N-body simulations and determined that if the LMC follows a decaying orbit, it could excite a warp of the required magnitude over a 6 Gyr timescale. Note that the N-body simulations of \citet{Garcia} yielded only a 25\% amplification of the total torque exerted by the LMC, rather than the required factor of five to reproduce the observed warp. \citet{Weinberg06} suggest that the discrepancy owed to an unlucky set of parameter choices by \citet{Garcia}. 

Altering the satellite's orbit changes the forcing frequency, and thereby affects both the amplitude and the orientation of the warp response.  \citet{Tsuchiya} adopted a MW model and orbital parameters such that the LMC's orbital period is $\sim$1.5 Gyr and the maximal apogalacticon distance is $\sim100-110$ kpc. But, for an isothermal sphere model, the K1 mean velocities imply an orbital period of $\sim3$ Gyr. The LMC thus completes only 2 orbits within 6 Gyr (Figure~\ref{fig:Isothermal}), rather than 4-4.5 orbits as in the \citet{Tsuchiya} model, and reaches a maximum apogalacticon distance of 250 kpc. 
If \citet{Garcia} found null warps owing to their halo parameter choices, the drastically different orbital values suggested by the new observations will surely limit the effectiveness of this mechanism. 

In the picture we have developed here, the LMC may be on its first passage about the MW.  Fly-by encounters have been considered in the context of halo wakes by \citet{Vesperini}. Here, the interloping galaxy induces a continuous spectrum of frequencies in the halo. Since only the low order modes are long-lived, the net effect is similar to that produced by a satellite on a quasi-periodic orbit. However, the magnitude of the effect is strongly dependent on the impact parameter and speed of the encounter: the disturbance is significant only after the perturbing galaxy has reached pericenter and the magnitude decreases as its velocity is increased.  Specifically, \citet{Vesperini} suggest that a perturber that is 5\% of the primary's mass and moving at $\sim$200 km/s could substantially perturb the primary's halo during a fly-by encounter.
 Since the new measurements imply that the LMC (mass $\sim$2\% of the MW) is traveling at $\sim400$km/s and we have concluded that the LMC is only currently at pericenter, it seems unlikely that the density perturbations induced by the LMC's passage through the MW halo can provide sufficient torque to excite a warp of the observed magnitude in the MW disk.



\subsubsection{Formation of the Magellanic Stream}
\label{subsubsec:stream}

The origin of the MS is highly controversial and different studies advocate for a variety of formation scenarios.  Historically, the MS has been thought of as a tidal feature, where gas is stripped from either the LMC or SMC at their last pericentric passage. But since tidal forces should work equally on both stars and gas, the lack of any stellar tidal feature associated with the MS \citep{NoStars} suggests that it may not have a purely tidal origin. Many works have also considered the role of hydrodynamic processes, such as ram pressure stripping. However, both of these mechanisms rely on the assumption that the L/SMC have undergone multiple pericentric passages about the MW. 

In the tidal stripping scenario, a satellite loses mass predominantly at pericentric passages \citep{Johnston95,Johnston98,Johnston99_2}. In the GN96 picture, the MS formed from material stripped from the SMC at its previous perigalactic approach 1.5 Gyr ago, which also coincided with a close encounter with the LMC. Our computed orbits ($\S$\ref{sec:results}) do not support this scenario: for the isothermal sphere model the previous pericentric passage occurred $\sim3$ Gyr ago; for our fiducial models we have determined that the L/SMC are currently at pericenter and only in the ``best case scenario'' ($\mu_W^\ast+4\sigma$) will another pericentric passage occur within a Hubble time. 
Without multiple pericentric passages, the effectiveness of tidal stripping is severely limited; 
along the fiducial orbits,
 the instantaneous tidal radius \citep{King} of the LMC(SMC) is greater than 15(7) kpc until roughly 0.2 Gyr ago, when the Clouds were $\sim$70 kpc away from the Galactic center. But the bulk of the HI in the LMC is distributed within a radius of $\sim$5 kpc \citep{Staveley}
 and the SMC has a line of sight depth of $\sim$7 kpc \citep{Harris04}. In addition, material outside the tidal radius is not stripped instantaneously \citep{Johnston98, Johnston99_2}. It is therefore doubtful that a tidal stripping scenario can produce the MS if the Clouds have only been subjected to significant tidal effects over the past 0.2 Gyr. Furthermore, the deviation of the projected orbits and the current position of the MS in the recent past presents a challenge to tidal stripping models, as discussed in $\S$\ref{subsec:MS}. 







During high velocity encounters, hydrodynamic interactions with the ambient medium are likely more significant than tidal distortions.
Specifically, various ram pressure stripping scenarios have been proposed to explain the formation of the MS, e.g., \citet[hereafter {\bf MD94}]{Moore}, HR94 and M05. Owing to the low density of the ambient MW halo gas ($n_e = 10^{-5}-10^{-4}$, see Table~\ref{Table:Constraints}) the \citet{Gunn} formalism for ram pressure stripping is insufficient to remove material instantaneously from the LMC disk.
 This is true even if the LMC's disk were face-on (which is not the case - the LMC disk is inclined $\sim$30 degrees from the line of motion) or if K1's new larger velocities were adopted. 

Alternatively, MD94 proposed that the MS was ram-pressure-stripped from the Clouds when they last passed through the extended ionized disk of the MW. They suggest this collision occurred $\sim$500 Myr ago at a galactocentric radius of 65 kpc. At this distance the density in the disk may be high enough for ram pressure to overcome the gravitational restoring force in the SMC or intracloud region. However, this scenario is not supported by our orbital analysis: for our computed orbits the LMC either never crosses the disk or does so at large radii ($>$400 kpc) and at much earlier times ($>$3 Gyr ago) (see Figures~\ref{fig:PMTimeLMC} and \ref{fig:PMDistLMC}).  Even for our high mass model the orbits do not support a disk crossing at radii $<$150 kpc (see Figure~\ref{fig:HighMassDist}).

Along a different line, M05 proposed a ``continuous stripping'' scenario, wherein the LMC gas disk edge is heated as a result of the compression exerted by the ambient halo gas. They argue that the corresponding increase in thermal energy can unbind the gas in the LMC disk on time scales comparable to the orbital time. This model implicitly assumes that the LMC has undergone multiple pericentric passages, which M05 achieved by using a significantly lower velocity than had been previously estimated (250 km/s).  For our orbits, the interaction time between the LMC and the halo gas ($<1-2$ Gyr) is likely too short for this to be a viable mechanism. Other hydrodynamic effects, such as Kelvin-Helmholtz and Rayleigh-Taylor instabilities (e.g. viscous stripping, \citet{Nulsen}), operate over even longer timescales and are expected to contribute minimally.

Alternate scenarios have been proposed where loosely bound material stirred by on-going tidal interactions within the LMC-SMC binary system is stripped from the Clouds or intracloud region (e.g. HR94 and \citet{Yoshizawa}). Such mechanisms may be viable, as they are less dependent on the assumption of multiple pericentric passages. We have also identified long-lived binary LMC-SMC states that are allowed by the data for all our computed orbits (see $\S$\ref{subsec:SMC}). However, such scenarios will still have difficulties explaining why the Clouds' orbits do not trace the MS ($\S$\ref{subsec:MS}). 

Our analysis suggests that the origin of the Magellanic stream remains an open question: all previously proposed mechanisms need to be revisited in light of the surprising consequences of the new proper motion measurements of K1 and K2.

As suggested by \citet{Fich}, the orbital history of the Clouds should be similar to that of the MS. Ultimately, a way to distinguish between formation mechanisms may thus come from the ability of the proposed mechanism to account for the deviation of the LMC's orbit with respect to the location of the MS on the sky (Figure~\ref{fig:MS}). Furthermore, it is also expected that the line of sight orbital velocities with respect to LSR ($v_{\rm LSR}$) should be roughly consistent with that of the HI observed in the stream today. 

\begin{figure}[t]
\begin{center}
\includegraphics[scale=0.43]{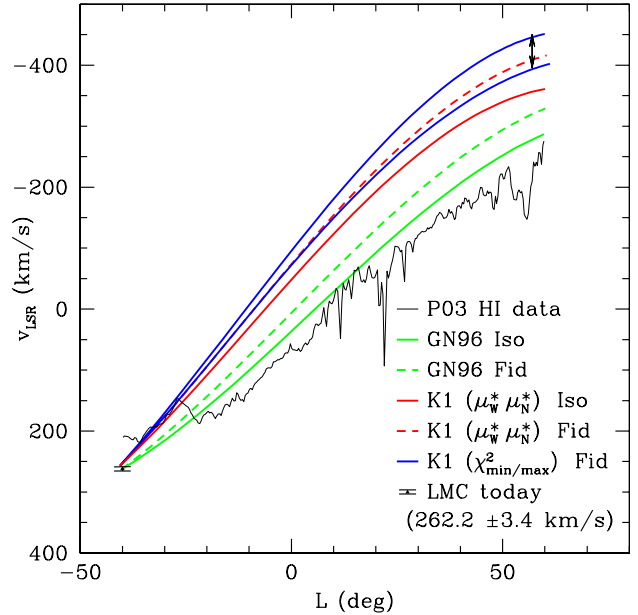}
\caption{Line of sight velocities with respect to LSR ($v_{\rm LSR}$) plotted as a function of Magellanic longitude (L) along the orbit shown in Figure~\ref{fig:MS}. The lines are color coded as in Figure~\ref{fig:MS}. The black line indicates the HI data of the MS from P03 (extending until MSV as defined by P03 in their figure 5. The $v_{\rm LSR}$ magnitudes from our orbit calculations are much higher than the P03 HI observations or the GN96 results (dashed green line). The black triangle indicates the current systemic LMC line of sight velocity (262.2 $\pm$ 3.4 km/s; vdM02). }
\label{fig:Vlsr}
\end{center}
\end{figure}

In Figure~\ref{fig:Vlsr}, the $v_{\rm LSR}$ is computed at each point
along the orbits depicted in Figure~\ref{fig:MS} and plotted as a
function of Magellanic longitude (L). The black line is the HI
velocity data of P03, estimated from the zeroth moment (intensity, $Int$) and
first moment (velocity) radio maps. At each Magellanic longitude (L),
the average over the Magellanic latitude (B) is plotted as $<V_{\rm
los}> = \sum V_{\rm los}(B) Int(B) dB/ \sum Int(B) dB$.  Once again,
the green line indicates the results for the isothermal GN96 model,
which approximates the HI velocity data.

The $v_{\rm LSR}$ values computed along the orbit corresponding to the mean K1 velocities in both the isothermal sphere (solid red line) and fiducial models (dashed red line; allowed values are bounded by the blue lines as indicated by the black arrow) are substantially higher than the observations of P03. 
  The tidal stripping model of \citet{Connors} is the only work to date to account for the recent \citet{Bruens2005} HI PARKES survey results; however, their orbital velocities are inconsistent with the new proper motions. 
The details of the predicted velocity structure of the MS are strongly dependent on the formation mechanism and further discussion is beyond the scope of this analysis. This will be the subject of a forthcoming paper.

\section{Conclusions}
\label{sec:conc}

This study has been motivated by two considerations: 1) previous works regarding the orbital evolution of the Magellanic Clouds have considered mainly isothermal sphere models for the MW. However, over the past decade our understanding of the structure and formation of dark matter halos has become more detailed and we can consider perhaps more realistic models to describe objects located at large Galactocentric radii.
2) Theoretical models for the MS have historically been used to constrain the proper motions of the Clouds - with the improved observations, we can now assume that the proper motions are known to some degree of accuracy. These two factors have very specific and surprising implications for our understanding of the orbital evolution of the Clouds. 


We have shown that for a $\Lambda$CDM-motivated model of the MW that is consistent with observational constraints and most theoretical models (such as model A$_1$ of KZS02; i.e., our fiducial model, $M_{vir} = 10^{12}\Msun$), proper motion estimates prior to the recent {\it HST} measurements result in drastically different orbital histories than previously predicted. Specifically, the GN96 estimate of the LMC's proper motion (based on theoretical models of the location and line of sight velocities of the MS) and the weighted average of pre-2002 ground-based and {\it Hipparcos} proper motion measurements (vdM02), both yield orbits where the LMC has completed only one passage about the MW within 10 Gyr and reached apogalacticon distances of 300-400 kpc, compared to the orbital period of 1.5 Gyr and apogalacticon distance of 100 kpc advocated by GN96 in an isothermal model. Thus, a drastic revision of our understanding of the orbital history of the Clouds is warranted in such a $\Lambda$CDM motivated picture for the MW. This result is independent of the proper motion measurements.

The new data give us additional information: the components of the
proper motion control very specific orbital parameters of the LMC. The
west component determines the tangential velocity and thereby sets
the number of pericentric passages (orbital period), apogalacticon,
location of disk crossing and the stability of the binary system.  The
north component dictates the location of the orbits with respect to
the current position of the MS on the sky. The implications of the new
data for the Magellanic system are discussed below, with specific
reference to the role of the individual proper motion components.

The {\it HST} proper motion measurements (K1) yield 3D velocities that are substantially higher ($\sim$100 km/s) than previously estimated, owing to the increase in the west component. This implies that the LMC is currently traveling at the local escape velocity. We subsequently
searched the error space within 4$\sigma$ of the mean values and found that the LMC could complete at most one orbit within a Hubble time, and only in cases where the west component of the proper motion was small ($|\mu_W^\ast|-4\sigma$). We tested whether our results were robust to changes in model parameters by considering a higher halo mass model ($M_{vir}=2\times10^{12}\Msun$) that was still consistent with observational constraints. Although the number of pericentric passages increased, the orbital period implied by the mean velocities is still longer than five Gyr. Perturbative effects owing to the presence of the SMC were shown to have a negligible effect on our orbital analysis owing to the LMC:SMC mass ratio of $\sim$10:1. Furthermore, our computed orbits do not preclude the existence of a stable binary LMC-SMC system, however this becomes more difficult to maintain as the MW mass increases. 

We conclude that either the Magellanic Clouds are on their first
passage about the MW or that the MW DM halo is well-modeled as an
isothermal sphere to distances $\gtrsim200$ kpc (i.e. substantially
more massive than $2\times10^{12}\Msun$, although note that this may
conflict with known observational constraints).  Even if the rotation
curve of the MW is flat out to distances of 200 kpc, the apogalacticon
distance and period of the LMC's orbit implied by K1's mean velocity
will have increased substantially compared to previous theoretical
estimates (220 kpc and 3 Gyr versus 100 kpc and 1.5-2 Gyr). This is a
direct result of the substantial increase in the $\mu_W$ proper motion
component measured by K1.

The north component of the proper motion does not affect the above
result, but as mentioned, it controls the location of the orbit when
projected on the plane of the sky. Using the mean K1 values, the LMC's
orbit deviates from the position of the MS by 7$\degr$ ($\sim7$ kpc)
on the sky. This result is {\it independent of both the $\mu_W$
component and the halo model} (see Figure~\ref{fig:MS}). This is
significant because while there are theoretical models that have
predicted tangential velocities as high as K1, they have all assumed
$\mu_N\sim 0$ (e.g., HR94 and LL82) - but this is not reconcilable
with the data. Even if we ignore the K1 measurement
($\mu_N=0.44\pm0.05$), the vdM02 weighted average of previous
measurements ($\mu_N = 0.34\pm0.16$) is not consistent with zero. This 
presents an impediment to both tidal and ram pressure stripping models of the 
MS, since both assume that the MS is co-located with the past orbit of the
Clouds.

In conclusion, if a $\Lambda$CDM-motivated model accurately describes the MW, and if the K1 results are correct, the Clouds are on their first passage about the MW. A first passage scenario is consistent with morphological studies of the satellite populations of the MW and M31, where the dIrr L/SMC appear as interlopers compared to the dSph galaxies that dominate the satellite population at small Galactocentric distances \citep{vandenbergh2006}, and is not statistically improbable from simulations of hierarchical structure formation. 
The origin of the LMC's high orbital angular momentum may also be better understood if the Clouds were not always orbiting about the MW. 

A first passage scenario has a number of unexpected phenomenological implications that warrant further investigation.  It supports a significant increase in the star formation rate of the Clouds during the past 1-3 Gyr, corresponding to when the Clouds first enter within the virial radius and begin to interact with the halo gas. It also drastically limits the timescale over which the Clouds and the MW can have interacted, making it highly unlikely that the LMC could have excited the warp in the MW disk. This forces a major reassessment of proposed formation mechanisms for the Magellanic stream, which depend strongly on the assumption that the Clouds have undergone multiple pericentric passages.  The implications for the formation of the MS are severe even if an isothermal sphere model of the MW is adopted: the orbital period and apogalacticon distances have increased, limiting the effectiveness of tidal and ram pressure stripping.

{\it Acknowledgements}
We would like to thank Beth Willman, Andrew Zentner and James Bullock for useful discussions on this project. We would also like to thank Mary Putman for providing us with her HI data of the MS. This research was partly supported by a post-graduate fellowship from the National Science and Engineering Research Council of Canada to GB and by NSF grant AST 03-07690 and NASA ATP grant NAG5-13292.

\bibliographystyle{apj}
\bibliography{Biblio}

\end{document}